%% file: arxiv.tex
\documentclass[9pt]{extarticle}
\usepackage[a4paper,left=18mm,right=18mm,top=17mm,bottom=18mm]{geometry}
\usepackage{amsmath,amssymb,mathtools}
\usepackage{graphicx}
\usepackage{xcolor}
\usepackage{booktabs}
\usepackage{placeins}
\usepackage[numbers,sort&compress]{natbib}
\usepackage[hidelinks]{hyperref}
\usepackage{microtype}
\usepackage{dblfloatfix}
\usepackage[T1]{fontenc}
\usepackage[utf8]{inputenc}

\newcommand{\nn}{\nonumber}
\newcommand{\be}{\begin{equation}}
\newcommand{\ee}{\end{equation}}
\newcommand{\bea}{\begin{eqnarray}}
\newcommand{\eea}{\end{eqnarray}}
\DeclareMathOperator{\Tr}{Tr}

\newcommand{\moy}[1]{\ensuremath{\left\langle #1 \right\rangle}}

\graphicspath{{figures/}}
\allowdisplaybreaks
\hypersetup{pdftitle={Typical number and stability of equilibria in large non-reciprocal complex systems},pdfauthor={Bertrand Lacroix-A-Chez-Toine, Pierre Le Doussal, Valentina Ros}}
\begin{document}
\twocolumn[{
\begin{center}
{\LARGE\bfseries Typical number and stability of equilibria in large non-reciprocal complex systems\par}
\vspace{7pt}
{\large Bertrand Lacroix-A-Chez-Toine$^{1}$, Pierre Le Doussal$^{2}$, and Valentina Ros$^{3}$\par}
\vspace{6pt}
{\small $^{1}$King's College London, Department of Mathematics, London WC2R 2LS, United Kingdom\par}
{\small $^{2}$Laboratoire de Physique de l'Ecole Normale Sup\'erieure, CNRS, ENS and PSL Universit\'e, Sorbonne Universit\'e, Universit\'e Paris Cit\'e, 24 rue Lhomond, 75005 Paris, France\par}
{\small $^{3}$Universit\'e Paris-Saclay, CNRS, LPTMS, 91405 Orsay, France\par}
{\small Corresponding author: \texttt{bertrand.lacroix\_a\_chez\_toine@kcl.ac.uk}\par}
\end{center}
\vspace{2pt}
\noindent\textbf{Abstract.} We consider a prototypical class of dynamics of $N\gg 1$ degrees of freedom subject to two competing terms, one favoring relaxation to a trivial equilibrium and one taking the form of a Gaussian force field promoting the emergence of many local equilibria. The latter is a non-gradient field, i.e. it includes a non-reciprocal component with a tunable strength. For this model, we compute the quenched complexity, i.e. the exponential growth rate of the typical number of equilibria, resolved by stability index. We compare it with the annealed complexity, i.e. the exponential growth rate of the average number of equilibria. We show that the two coincide only in the fully non-reciprocal limit: for any finite degree of reciprocity, typical realizations contain exponentially fewer equilibria than predicted by the average count. Increasing reciprocity enhances the discrepancy between the two. The quenched complexity depends explicitly on the covariance of the random force field and on the degree of non-reciprocity: the universality observed at the annealed level breaks down. The average counting of equilibria correctly identifies the location of the topology trivialization transition, but it does not capture the typical organization of equilibria near this transition, nor the stability index of the dominant equilibria.\par
\vspace{7pt}
}]
\input{main_text.tex}
\clearpage
\onecolumn
\appendix
\setcounter{section}{0}
\renewcommand{\thesection}{S\arabic{section}}
\setcounter{subsection}{0}
\renewcommand{\thesubsection}{\thesection.\arabic{subsection}}
\setcounter{equation}{0}
\renewcommand{\theequation}{S\arabic{equation}}
\setcounter{figure}{0}
\renewcommand{\thefigure}{S\arabic{figure}}
\setcounter{table}{0}
\renewcommand{\thetable}{S\arabic{table}}
\renewcommand{\theHsection}{S\arabic{section}}
\renewcommand{\theHsubsection}{S\arabic{section}.\arabic{subsection}}
\renewcommand{\theHequation}{S\arabic{equation}}
\renewcommand{\theHfigure}{S\arabic{figure}}
\renewcommand{\theHtable}{S\arabic{table}}
\section*{Supplementary Information}
\input{supplement.tex}
\clearpage
\bibliographystyle{unsrtnat}
\bibliography{references}
\end{document}

%% file: main_text.tex
Many rapidly developing areas of science require theoretical tools to describe out-of-equilibrium dynamics in high-dimensional systems made of many coexisting and interacting units. Examples include ecological communities \cite{nguyen2025inferring,amor2020transient, venturelli2018deciphering},
biological and artificial neural circuits \cite{vyas2020computation,international2025brain, shenoy2021measurement}, or networks of agents \cite{scholl2021market, asano2021emergent}. 
In most of these settings, the units interact with high connectivity,
and the interactions tend to be non-reciprocal (asymmetric), reflecting the fact that the influence of one component on another is not mirrored in return. Moreover, interactions are typically heterogeneous, both in strength and sign. Following a longstanding statistical physics approach~\cite{amari1971characteristics, may1972will, follmer1974random, sherrington1975solvable}, it is natural to model these couplings with randomness, thereby defining \emph{ensembles} of systems. The aim is then to identify features that are robust at the ensemble level, and do not depend on microscopic details of a particular realization of the interaction couplings. 

Randomness provides an efficient way to capture a central feature of complex systems, namely that very different configurations of a system can give rise to similar macroscopic behavior or performances. One manifestation of this is the possible coexistence of multiple fixed points (or equilibria) of the system of equations governing the dynamics. 
Depending on the context, these fixed points may represent ecological equilibria, stationary neural activity patterns, local optima in learning, or metastable configurations in disordered materials. Counting them, and characterizing their statistical properties~\cite{
bray1980metastable,
fyodorov2007density, bray2007statistics, fyodorov2016topology, 
ros2020distribution,
fyodorov2020manifolds, 
gershenzon2023site, 
kent2024arrangement, pacco2025triplets, fyodorov2025superposition} is a crucial step toward understanding the range of dynamical behaviors available to the system. Indeed, rich out-of-equilibrium dynamics often occurs alongside the coexistence of many equilibria~\cite{
kadmon2015transition,
folena2020rethinking, pearce2020stabilization, arnoulx2024many, fournier2026nonreciprocal, fournier2026high}.
This is especially interesting for non-reciprocal interactions, where the dynamics is in general not gradient-like, 

\noindent and fixed points are not simply minima or saddles of an underlying potential landscape~\cite{ros2023high}.
In random high-dimensional systems, however, even the question “how many fixed points are there?” does not have a unique answer. One may compute the ensemble-{\it averaged} number, or ask how many fixed points occur in \emph{typical} (i.e., highly probable) realizations of the disorder. While in low-dimensional settings this distinction may be mild, in high dimension it can become decisive, as the number of fixed points can grow exponentially with the dimension.  Averages of this exponentially growing quantity over the whole ensemble can be dominated by exponentially rare realizations with anomalously many fixed points, rather than by representative realizations. In this sense, averages may describe systems that almost never occur.
Deriving statistics that are not dominated by rare events is a longstanding challenge in the physics of disordered systems, which has motivated the development of dedicated techniques~\cite{edwards1975theory, mezard1988spin, parisi_urbani_zamponi2020book}.

In this work, we consider a prototypical model~\cite{fyodorov2016nonlinear,ben2021counting,lacroix2022counting} of high-dimensional, random  equations describing the dynamics of a system with tunable non-reciprocity. This model is deliberately generic: it does not aim to reproduce the microscopic structure of specific ecological or neural systems, or systems of interacting agents; rather, it isolates three ingredients common to many of them: high dimensionality, heterogeneous interactions, and tunable reciprocity. We compute the exponential growth rate of the number of fixed points in typical realizations of the disorder and compare it with the previously known growth rate of the ensemble average~\cite{fyodorov2016nonlinear,ben2021counting,lacroix2022counting}. This comparison shows that average counting correctly locates the onset of an exponential number of equilibria, but it gives a biased description of the fixed-point population itself. While the growth rate determined by the average counting is universal, the typical one reveals a dependence on the statistics of the random force field governing the system's dynamics, and in particular on the degree of reciprocity. Only in the fully non-reciprocal limit does the annealed count become representative. Rare-sample bias appears as soon as some degree of reciprocity is introduced, and the discrepancy between typical and average counting becomes stronger as reciprocity is increased. 

\begin{figure}[t!]
\includegraphics[width=.98\linewidth]{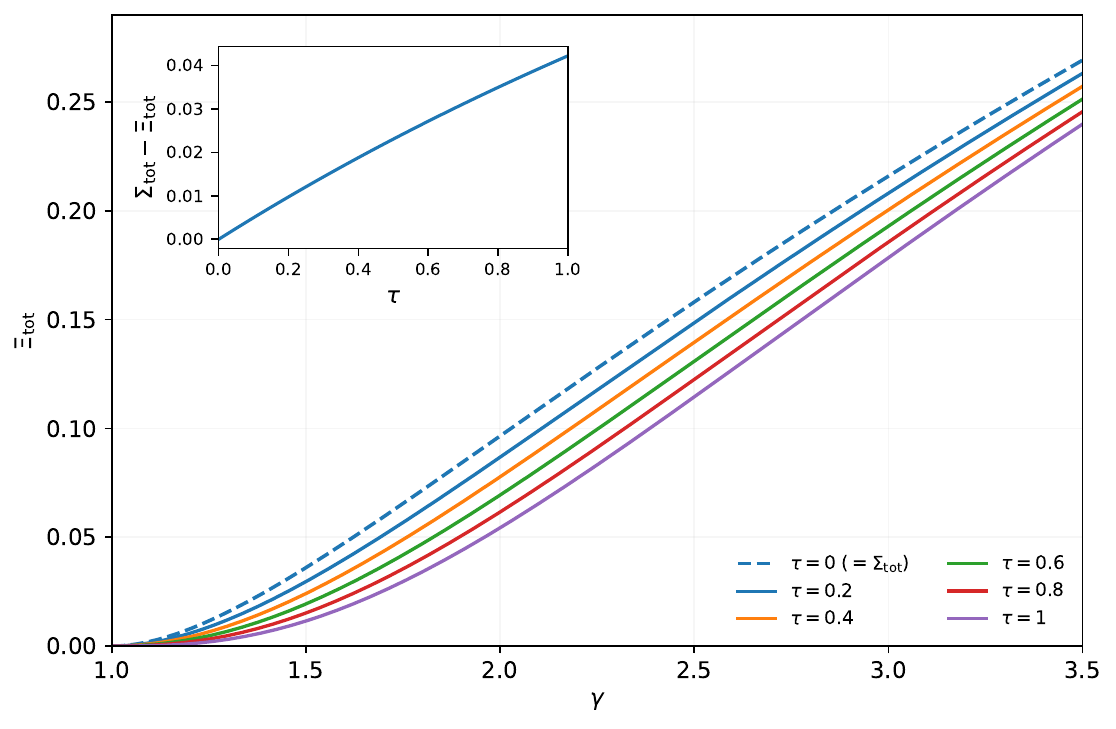}
\caption{Quenched complexity $\Xi_{\rm tot}$ as a function of the rescaled disorder $\gamma$, for different values of the reciprocity $\tau$. The dashed light blue line is the annealed complexity $\Sigma_{\rm tot}$, which coincides with the quenched one only for $\tau=0$. \emph{Inset.} Difference between annealed and quenched complexities as a function of $\tau$
for $\gamma=2$. Both plots correspond to $\Gamma_V(q)=\tau\Gamma(q)$ and $\Gamma_A(q)=(1-\tau)\Gamma(q)$ with $\Gamma(q)=\exp(-q)$.}
\label{CompBeh}
\end{figure}

\section*{Fixed points and quenched complexity}
We consider a model of  $N \gg 1$ coupled, autonomous ordinary differential equations of the generic form 
\begin{equation}
\partial_{t}{\bf x}={\bf f}({\bf x})\,,\,\,f_i({\bf x})=-\mu_i x_i+F_i({\bf x})\,,\,\,i=1,\cdots,N\,,
\label{dyn}    
\end{equation}
where ${\bf x}\in \mathbb{R}^N$, the relaxation rates $\mu_i$'s are positive and ${\bf F}({\bf x})$ is a random force field. We assume the force to be non-gradient, meaning that there is no scalar potential $ V({\bf x})$ such that ${\bf F}({\bf x})=-\nabla V({\bf x})$. This model describes the dynamics of a large collection of units subject to two competing contributions: a linear one-body term which drives each unit exponentially toward a threshold value (set to zero in our normalization), and a nonlinear interaction term $F_i({\bf x})$, which couples all units in a heterogeneous way and encodes their random interactions. Competition between these two terms leads to transitions in the population of equilibria ${\bf x}_{*}$, that are the configurations satisfying ${\bf f}({\bf x}_*)={\bf 0}$. For ${\bf F}=0$, the system trivially admits the unique fixed point ${\bf x}_*=0$; the non-linear term, if strong enough, can generate a multiplicity of them. We stress that, while for gradient fields ${\bf F}({\bf x})=-\nabla V({\bf x})$ the dynamics can only converge in the long-time limit to a fixed point configuration, for non-gradient fields it may asymptotically reach e.g. a fixed point, a limit cycle or a chaotic attractor~\cite{sompolinsky1988chaos,pearce2020stabilization,  
fournier2026chaos}. We focus on fixed points because their statistical properties can provide insight into the structure of the dynamics, even in regimes where chaotic behavior emerges~\cite{fournier2026nonreciprocal, stubenrauch2025fixed, wainrib2013topological, yang2025relationship}.

Relevant information for the system's dynamics is the linear stability of fixed points, which  is governed by the Jacobian matrix
\be
{\cal J}_{ij}({\bf x}_*)=\left.\frac{\partial f_i}{\partial x_j}\right|_{{\bf x}_*}=\left.\frac{\partial F_i}{\partial x_j}\right|_{{\bf x}_*}-\mu_i\delta_{ij},
\ee
where $\delta_{ij}=1$ if $i=j$ and $0$ otherwise is the Kronecker delta. This matrix is obtained 
 by linearizing \eqref{dyn} at a fixed point ${\bf x}_*$. The instability index $k({\bf x}_*)$ is the number of eigenvalues of ${\cal J}({\bf x}_*)$ with positive real part. 
 
 Two types of question may be considered: local questions that concern a particular fixed point ${\bf x}_*$, and global questions that concern all the different fixed points of the dynamics. A standard local question, first asked by May in his seminal paper \cite{may1972will}, is under what conditions is a {\it given} equilibrium stable? A global question, partly investigated in \cite{fyodorov2016nonlinear,ben2021counting}, is instead how many equilibria does the dynamics display and what is the distribution of instability indices among these? The randomness in the force field requires us to specify in which sense the count of equilibria should be understood, since the number of equilibria is itself a random variable. The simplest statistic, the average number $\moy{{\cal N}_{N,k}}=\sum_{n\geq 0}n\,{\rm Prob}.\left[{\cal N}_{N,k}=n\right]$ of equilibria with instability index $k$, was investigated in \cite{fyodorov2016nonlinear,ben2021counting,lacroix2022counting} for a class of models belonging to the family (\ref{dyn}). The analysis revealed regimes of parameters in which $\moy{{\cal N}_{N,k}}$ grows exponentially with the dimension $N$. In general random variables exhibiting such fast growth need not concentrate, implying that their average and typical values do not coincide, even asymptotically. 

When the number of equilibria is exponentially large with the dimension,
 the quantity of interest is then its typical value ${\cal N}_{{\rm typ},k}\sim e^{\moy{\ln{\cal N}_{N,k}}}$, observed in almost any realization of the random system when $N \gg 1$. Asymptotically, the average and typical number of fixed points are characterised respectively by the ``annealed" $\Sigma(\kappa)$ and ``quenched" complexity  $\Xi(\kappa)$ of equilibria with extensive index $k=N\kappa$, defined as
\be
\Sigma(\kappa)=\lim_{N\to \infty}\frac{\ln \moy{{\cal N}_{N,N\kappa}}}{N}, \; \Xi(\kappa)=\lim_{N\to \infty}\frac{\moy{\ln \left({\cal N}_{N,N\kappa}+1\right)}}{N}\,.\label{def-comp}
\ee
Note that we consider the logarithm of one plus the number of equilibria to avoid divergences if the number of equilibria is zero, which happens with non-zero probability at finite $N$. When the complexities are non-negative, $\Sigma(\kappa) \geq \Xi(\kappa)\geq 0$; the bound is a simple consequence of Jensen's inequality, and reflects the fact that the average can be skewed by exponentially rare realizations of the disorder that display an exponentially larger number of equilibria than typical realizations, see Fig.~\ref{CompBeh}. As the total number of equilibria satisfies ${\cal N}_{N,\rm tot}=\sum_{k=0}^N {\cal N}_{N,k}\geq 1$, the total (annealed and quenched) complexities can be obtained by maximizing with respect to $\kappa$ in the expression above. We present results for the value of the quenched complexities $\Xi(\kappa)$ for $0\leq \kappa\leq 1$ when this rate is positive and for the boundary curve where this rate vanishes. We focus on the same choice of statistics of the random forces considered in \cite{fyodorov2016nonlinear,ben2021counting,lacroix2022counting}, where the annealed complexities $\Sigma(\kappa)$ were computed in detail. In the following, we refer to equilibria with $\kappa=0$ as stable, although our analysis only resolves the index on the extensive scale: $\kappa=0$ corresponds to equilibria with $k=o(N)$ unstable directions, which includes but is not restricted to strictly stable ones ($k=0$).

\section*{Model assumptions and previous results}
We consider the model of non-gradient Gaussian field introduced in \cite{fyodorov2016nonlinear}, which can be expressed as the sum of a conservative field (gradient term) and a divergence-free term
\begin{equation}\label{eq:ForceDec}
   F_i({\bf x})=-\frac{\partial V ({\bf x})}{\partial x_i}+\frac{1}{\sqrt{N}}\sum_{j=1}^N \frac{\partial A_{ij}({\bf x})}{\partial x_j}\,.
\end{equation}
The potential $V({\bf x})$ and the  antisymmetric matrix field $A({\bf x})$ are independent, Gaussian, and they both have zero average entries. The covariances satisfy
\begin{align}
    \moy{V({\bf x}_a)V({\bf x}_b)}&=\Gamma_V\left(|{\bf x}_a-{\bf x}_b|^2\right)\,,\\
    \moy{A_{ij}({\bf x}_a)A_{kl}({\bf x}_b)}&=\left(\delta_{ik}\delta_{jl}-\delta_{il}\delta_{jk}\right)\Gamma_A\left(|{\bf x}_a-{\bf x}_b|^2\right)\,,
\end{align}
where $\Gamma_{A,V}(q)=\int_0^{\infty}d\sigma_{A,V}(s)e^{-s q}$ for some positive finite measure $d\sigma_{A,V}(s)\geq 0$ on $\mathbb{R}^+$ \cite{schoenberg1938metric}. We assume additionally that it has a finite fourth moment $\Gamma_{A,V}^{(4)}(0)=\int_0^{\infty}s^4 d\sigma_{A,V}(s)<\infty$, with both $\Gamma_{A,V}''(q)\geq 0$ and $\Gamma_{A,V}'(q)\leq 0$ for any $q\in[0,\infty)$.  
The average $\moy{\cdot}$ is with respect to realizations of both $V$ and $A$. We consider the case of isotropic relaxation rates $\mu_i=\mu$. In this case, the  two main parameters in this model are given by the rescaled disorder parameter $\gamma\geq 0$ and the degree of symmetry of the field (or equivalently of reciprocity of interactions) $0\leq \tau\leq 1$ defined as
\be
\gamma=\frac{4N(\Gamma_A''(0)+\Gamma_V''(0))}{\mu^2}\,,\,\,\tau=\frac{\Gamma_V''(0)}{\Gamma_A''(0)+\Gamma_V''(0)}\,.
\ee
In particular, for $\tau=0$ the field is divergence-free (fully non-reciprocal interactions) while for $\tau=1$ the field is gradient (reciprocal). For simplicity and conciseness, we only display in the main text the results in the case where the correlation functions $\Gamma_V(q)=\tau\Gamma(q)$ and $\Gamma_A(q)=(1-\tau)\Gamma(q)$. 
The case of distinct correlation functions and anisotropic rates $\mu_i$ can however be treated explicitly (see the SI). 

It is useful to recall the main results obtained for the annealed complexity~\cite{fyodorov2016nonlinear,ben2021counting,lacroix2022counting}. The total annealed complexity was found to be (with $\gamma=1/m^2$ from \cite{fyodorov2016nonlinear})
\be
\Sigma_{\rm tot}=\frac{1}{2}\left(\frac{1}{\gamma}-1+\ln \gamma\right)\,,\,\,\gamma\geq 1\,,\label{tot_ann}
\ee
and $\Sigma_{\rm tot}=0$ otherwise. Two observations follow from this expression. First, the system undergoes a trivialization transition~\cite{fyodorov2004complexity} at the critical disorder strength $\gamma_c=1$, separating a {\it trivial} phase with a single equilibrium \cite{ben2021counting} (corresponding to vanishing complexity) from a {\it complex} phase with an exponentially large average number of equilibria. Analogous transitions arise in a variety of models~\cite{sompolinsky1988chaos, opper1992phase, yoshino2007statistical} displaying the same underlying competition, schematized in model (\ref{dyn}), between stabilizing one-body terms that admit a unique fixed point and random nonlinear terms that generate complexity. They have a dynamical counterpart: in the trivial phase, trajectories are expected to converge to the unique stable fixed point of the dynamical equations, while in the complex phase the system typically displays a persistent out-of-equilibrium dynamics (away from the gradient limit). The second observation is that strikingly, to leading exponential order the average number of fixed points is universal: it does not depend on the symmetry index~$\tau$, nor on the function $\Gamma(q)$ characterizing the statistics of the forces. 

\begin{figure*}[t!]
\includegraphics[width=.48\linewidth]{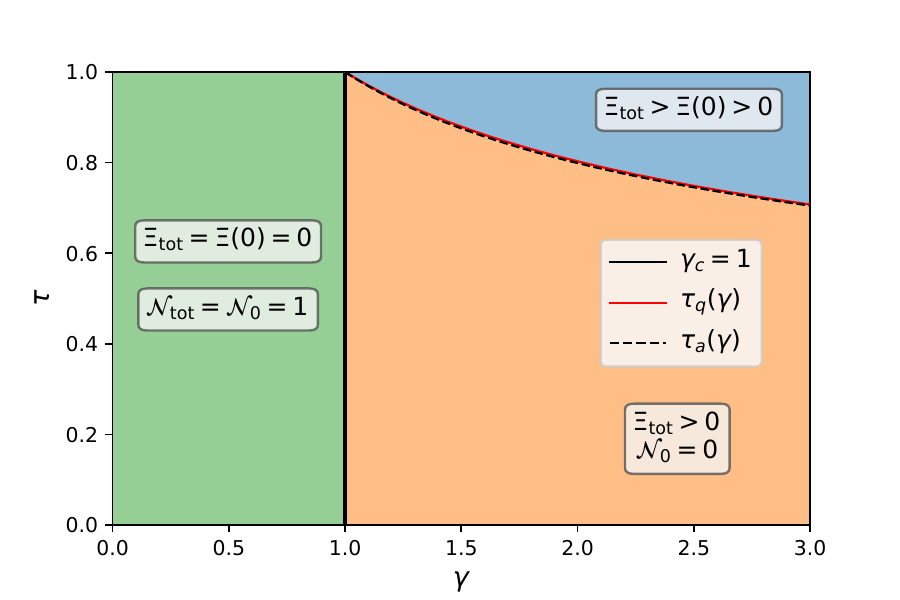}
\includegraphics[width=.46\linewidth]{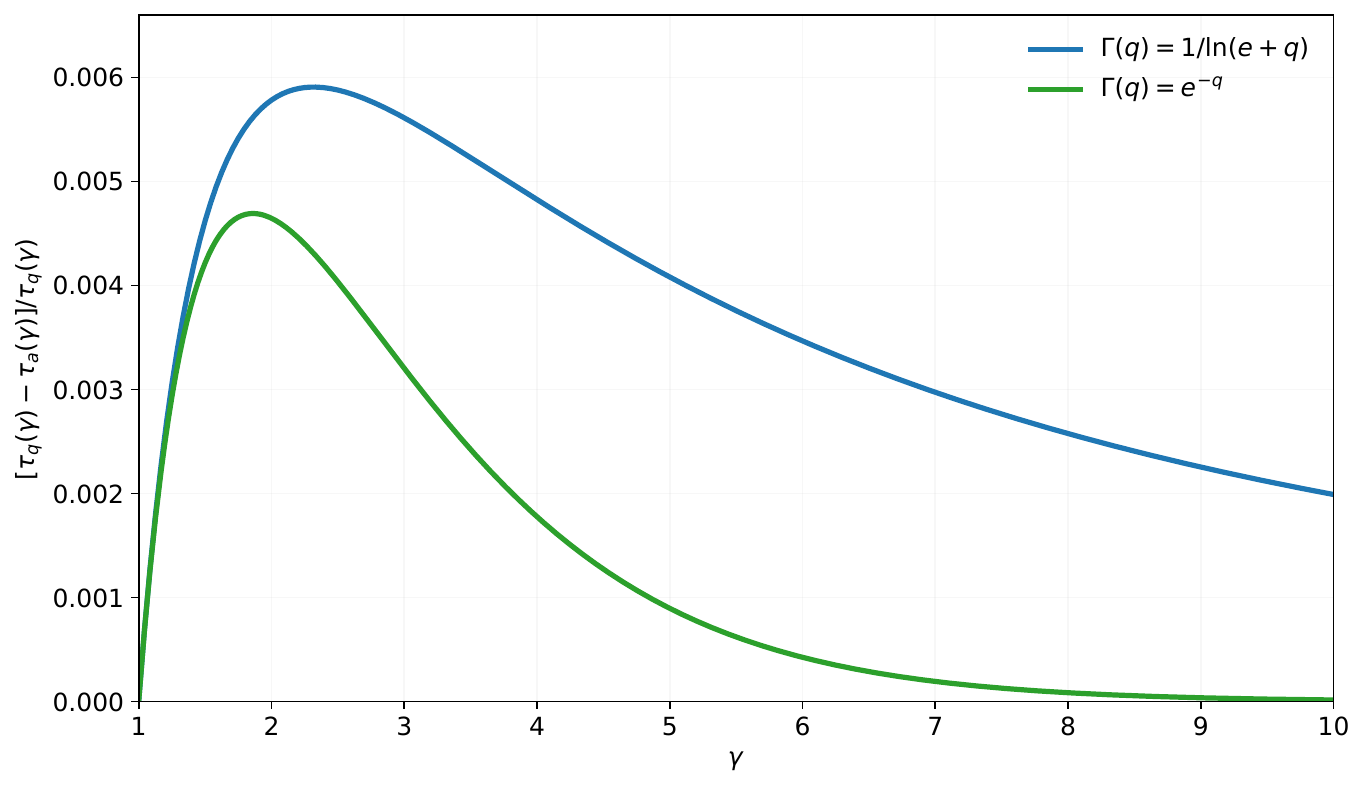}
\caption{\emph{Left.} Complexity phase diagram of the model with $\Gamma(x)=e^{-x}$ and uniform $\mu_i$. The vertical line corresponds to the topology trivialization transition. In the trivial phase $\gamma<1$, the number of total and strictly stable equilibria ${\cal N}_{\rm tot}={\cal N}_0$ is one with probability tending to one and both complexities vanish. The dashed black line (annealed) and red line (quenched) correspond to $\tau_a(\gamma)$ in \eqref{eq:TauCAnn} and $\tau_q(\gamma)=\tau_{-}(\gamma,0)$ in \eqref{tau_pm}, respectively. These are the lines along which the complexity of fixed points with $\kappa=0$ vanishes. The probability to have exactly zero stable equilibria ${\rm Prob}.\left[{\cal N}_0=0\right]$ goes to one below the line $\tau_a(\gamma)$ while it goes to zero above the line $\tau_q(\gamma)$. In between the two lines $\tau_a(\gamma)<\tau<\tau_q(\gamma)$ we cannot conclude on its value from our analysis. \emph{Right.} Relative shift of the transition lines for different choices of covariance $\Gamma(x)$.  }
\label{phasediag}
\end{figure*}

The annealed complexity of equilibria with fixed index reads
\be
\Sigma(\kappa)=\Sigma_{\rm tot}-\frac{1+\tau}{2\gamma\tau}(\sqrt{\gamma}z(\kappa)-1)^2\,,\,\,\kappa>0\,,
\ee
where $z(\kappa)$ is defined by
\be
\kappa=\frac{1}{\pi}\left[{\rm arccos}\left(z(\kappa)\right)-z(\kappa)\sqrt{1-z(\kappa)^2}\right]\;.\label{kappa_z_eq}
\ee
For $\kappa=0$ (stable fixed points), $z(0)=1$ and the expression above is only valid for $\gamma>1$ while $\Sigma(0)=\Sigma_{\rm tot}=0$ for $0\leq \gamma\leq 1$. More precisely, it was shown in \cite{ben2021counting} that in the trivial phase ($\gamma<1$), there is a unique stable equilibrium ${\cal N}_{\rm tot}={\cal N}_0=1$ with probability tending to one in the limit $N\to \infty$. Similarly, in regimes where $\Sigma(\kappa)<0$, the number of equilibria with corresponding stability ${\cal N}_{N,N\kappa}$ is zero  with probability tending to one as $N\to \infty$, and there is typically no such equilibrium. From this expression, it is simple to realize that the highest number of equilibria have index $\kappa_{a}(\gamma)=0$ in the trivial phase $0\leq \gamma\leq 1$, and  $z(\kappa_{ a}(\gamma))=\gamma^{-1/2}$ for fixed $\gamma>1$ and irrespective of the value of $\tau$. In the complex phase, on average the most numerous fixed points are dynamically unstable ($\kappa_a>0$). Within this phase, depending on the values of $\gamma, \tau$, the average number of stable fixed points is either exponentially large  ($\Sigma(0)>0$), or exponentially small  ($\Sigma(0)<0$), signaling 
an  exponentially small probability to have any stable equilibrium. The transition between these two phases is predicted to occur for $\gamma>1$ at the value
\be\label{eq:TauCAnn}
\tau_a(\gamma)=\frac{(\sqrt{\gamma}-1)^2}{2\sqrt{\gamma}-2\gamma+\gamma\ln\gamma}\;.
\ee
A complexity phase diagram accounting for  the different regimes is given in Fig.~\ref{phasediag}.

\section*{The typical number of fixed points}
The quenched calculation of the complexity allows one to test which features of the annealed phase diagram are typical, and which instead come from rare realizations with anomalously many equilibria that dominate the average.
We compute the quenched complexity by means of the replicated Kac-Rice formalism \cite{ros2019complex}. This combines methods from random matrix theory, already needed in the annealed calculation \cite{fyodorov2004complexity, AufBenArusCerny2013}, with the replica formalism, which is required to extract the contribution of typical realizations of the randomness. This method requires  an ansatz on the structure of the correlations between different equilibria of the dynamical equations: we present results obtained under the replica-symmetric ansatz, in which all pairs of equilibria are assumed to be statistically equivalent and characterized by the same mutual correlation.  
We show that this ansatz is stable (with respect to continuous replica symmetry breaking) for the covariances and regimes of parameters considered in this work, following \cite{AT78} (see {\bf Materials and Methods} for details). 

We find that the total quenched complexity is given by the explicit expression
\be
\Xi_{\rm tot}
=\Sigma_{\rm tot} -\frac{\tau}{2}\frac{ (\gamma -1)^2   v(\gamma)}{\gamma [ 1+  \tau  v(\gamma)]}\,,\,\,\;\gamma\geq 1\,,
\label{quenched_tot}
\ee
and $\Xi_{\rm tot}=0$ otherwise.
The function $v(\gamma)$ is defined as
\be
v(\gamma)=\frac{\Gamma''\left({\cal Q}(\gamma)\right)}{\Gamma''(0)}\,,\,\,{\rm where}\,\,\gamma=\frac{{\cal Q}(\gamma)\Gamma''(0)}{\Gamma'({\cal Q}(\gamma))-\Gamma'(0)}\,.
\ee
The second equation defines ${\cal Q}(\gamma)$, that is interpretable as the squared distance between equilibria in a typical realization of the disorder. Notice that  $\Gamma(q)$ being completely monotone implies that $\Gamma''(q)$ is a positive and decreasing function, and thus that $0\leq v(\gamma) \leq 1$. By concavity of $\Gamma'(q)$, the function ${\cal Q}(\gamma)$ is an increasing function of $\gamma$ and $v(\gamma)$ is a decreasing function of $\gamma$ with $v(1)=1$ and $v(\gamma)\to 0$ as $\gamma\to \infty$. The sign of $v(\gamma)$ ensures that $\Xi_{\rm tot} \leq \Sigma_{\rm tot}$, as expected. The inequality is saturated for $\gamma\to 1_+$, where the two complexities both vanish, identifying the trivialization transition. The 
quenched complexity vanishes quadratically at  the transition ($\gamma\to 1_+$) for any $0\leq \tau<1$ while for a gradient dynamics $\tau=1$, it vanishes cubically
\begin{align}
\Xi_{\rm tot}=&\frac{1-\tau}{4(1+\tau)}\left(\gamma-1\right)^2\label{qu_exp}\\
&+\frac{(\tau(5+\tau)-2)}{6(1+\tau)^2}\left(\gamma-1\right)^3+O\left(\gamma-1\right)^4.  \nn 
\end{align} 

Next, we consider the quenched complexity with a fixed fraction $\kappa$ of unstable directions. While the total complexity is always non-negative by construction, this is not the case for the complexity at fixed fraction $\kappa$. We obtain the quenched complexity through a proxy $\hat \Xi(\kappa)$ which coincides with $\Xi(\kappa)$ whenever it is positive, and vanishes on the same line(s) (see {\bf Materials and Methods} for details). Our expression for the quenched complexity is thus valid
in the complex phase $\gamma\geq 1$ and in a finite range of values of $\tau$, i.e.
\begin{align}
\tau_-(\gamma&,\kappa)\leq \tau\leq \min(1,\tau_+(\gamma,\kappa))\;,\\
\tau_\pm(\gamma&,\kappa)=\frac{1}{2}\left(-b\pm\sqrt{b^2-\frac{4(\sqrt{\gamma}z(\kappa)-1)^2}{\gamma v(\gamma)(\sqrt{\gamma}-z(\kappa))^2}}\right)\;,\label{tau_pm}\\
    b=&\frac{(1+v(\gamma))(1+z(\kappa)^2)-(1-v(\gamma))\ln \gamma}{(z(\kappa)-\sqrt{\gamma})^2 v(\gamma)}\\
    &-2z(\kappa)\frac{(1+\gamma v(\gamma))}{(z(\kappa)-\sqrt{\gamma})^2 \sqrt{\gamma}v(\gamma)}\;,\nn
\end{align}
where $z(\kappa)$ is obtained by inverting \eqref{kappa_z_eq}.
In this range, the quenched complexity at fixed $\kappa$ reads 
\begin{align}
\Xi(\kappa)&=\frac{v(\gamma)(\sqrt{\gamma}-z(\kappa))^2}{2\tau(1-v(\gamma))}(\tau-\tau_-(\gamma,\kappa))(\tau_+(\gamma,\kappa)-\tau)\label{quenched_kappa}\\
    &=\Sigma(\kappa)- \frac{v(\gamma) \left[\gamma  \tau -\sqrt{\gamma } (\tau +1) z(\kappa)+1\right]^2}{2 \gamma  \tau  (1-v(\gamma))}.\nn
\end{align}
Once more, $0\leq v(\gamma)\leq 1$ guarantees that $\Xi(\kappa) \leq \Sigma(\kappa)$. 

\begin{figure}[t!]
\centering
\includegraphics[width=.9\linewidth]{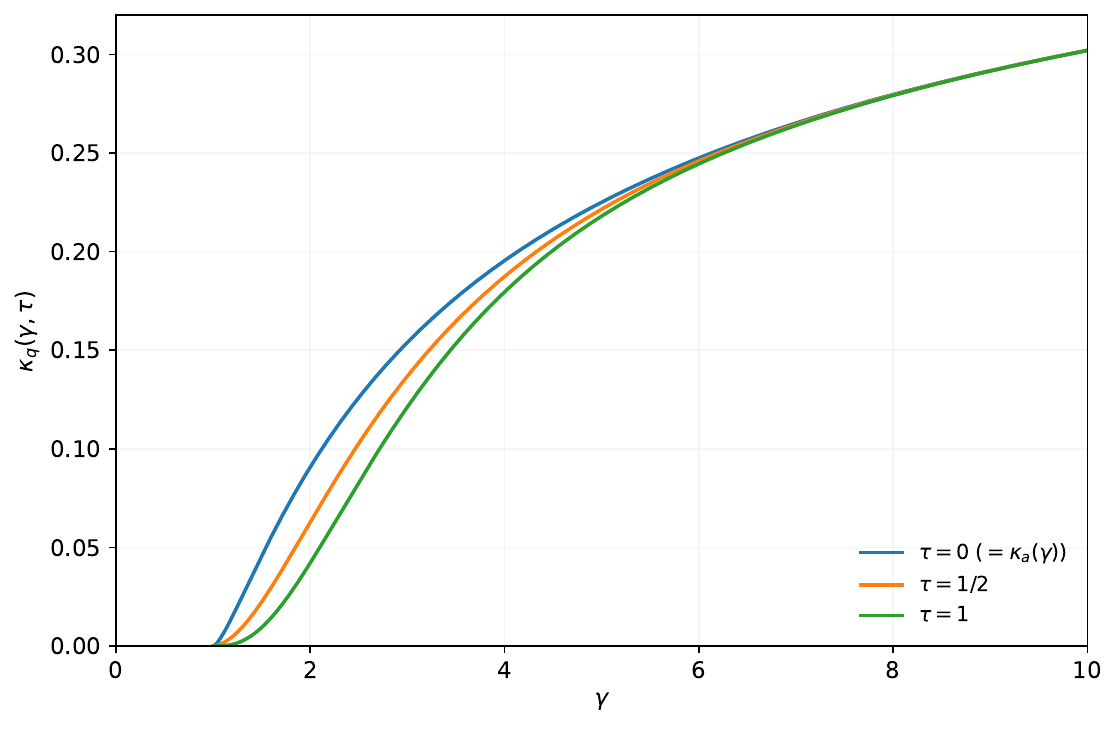}
\caption{Plot of the index $\kappa_q(\gamma,\tau)$ of dominant (most numerous) equilibria as a function of $\gamma$ for the covariance $\Gamma(q)=e^{-q}$ and for $\tau=0,1/2,1$ respectively in blue, orange and green. For $\tau=0$, the value of $\kappa_q(\gamma,0)\equiv \kappa_a(\gamma)$ is the same as the value obtained from the annealed complexity. The index decreases as a function of $\tau$, indicating that more reciprocal interactions favor more stable equilibria.}
\label{kappa-star-fig}
\end{figure}

The expression for $\Xi_{\rm tot}$ is recovered by optimising with respect to $\kappa$ (or equivalently with respect to $z(\kappa)$). This gives us the information on the stability of the equilibria dominating at arbitrary $\gamma$. The index $\kappa_q\equiv\kappa_q(\gamma,\tau)$ with the largest typical number of equilibria $\Xi_{\rm tot}=\max_{\kappa}\Xi(\kappa)=\Xi(\kappa_q)$ is equal to zero in the trivial phase, while in the complex phase it satisfies
\be \label{eq:IndexQ}
z(\kappa_q)=\frac{\displaystyle 1+\tau\gamma v\left(\gamma\right)}{\displaystyle\sqrt{\gamma}\left(1+\tau v\left(\gamma\right)\right)}\;.
\ee
Plots of $\kappa_q$ for different values of $\tau$ are given in Fig.~\ref{kappa-star-fig}. Even though stable fixed points $\kappa=0$ are never dominant for $\gamma>1$, one finds an exponentially large number of them (in typical instances of the system) whenever   $\tau>\tau_q(\gamma)=\tau_-(\gamma,0)$ given in \eqref{tau_pm}. As shown in Fig.~\ref{phasediag}, for this model different $\Gamma(q)$ give transition lines $\tau_q(\gamma)$ that are shifted by small relative amounts: they lie close to one another, and hence also close to the line in \eqref{eq:TauCAnn} obtained from the annealed calculation.

\subsection*{Divergence-free field $\tau=0$}
In the special case where $\tau=0$, one can check that (with our definition in \eqref{def-comp}) the complexity $\Xi(\kappa)=0$ for any value of index $\kappa\neq \kappa_q$ while $\Xi(\kappa_q)=\Xi_{\rm tot}\geq 0$. Thus the probability of having equilibria with an index $\kappa\neq \kappa_q(\gamma,0)$ different from the typical one decays more rapidly than $e^{-O(N)}$. This was already observed for the annealed complexity in \cite{ben2021counting}. In addition, inserting $\tau=0$ in \eqref{quenched_tot}, one immediately sees that the quenched and annealed complexity coincide for any $\gamma\geq 0$. As shown in SI, this result can be further refined by computing the factorial moments of the total number of equilibria. In the high-dimensional limit, they satisfy the simple identity
\be
\lim_{N\to \infty}\frac{1}{N}\ln\moy{\prod_{p=0}^{n-1}\left({\cal N}_{N,\rm tot}-p\right)}=n\Xi_{\rm tot}\;.\label{div-free}
\ee
Thus, for $\tau=0$, the factorial moments of the total number of equilibria factorize at leading exponential order.

\section*{Discussion}
Our results show that rare disorder realizations bias the ensemble-average counting of fixed points, so that annealed calculations capture only part of the fixed-point phenomenology. The clearest signature of this is a loss of universality. Comparing the quenched complexity in~\eqref{quenched_tot} with its annealed counterpart in~\eqref{tot_ann} shows that, while the average growth rate depends only on the rescaled disorder $\gamma$, the typical growth rate depends also on the degree of symmetry $\tau$ and on the covariance function $\Gamma(q)$, through the function $v(\gamma)$. Thus, the universality of the annealed result does not extend to typical realizations, and the quenched complexity contains finer information about the structure of the dynamics. This reflects the fact that, unlike the average count, the typical number of fixed points is sensitive to the correlations between them, which are inherited from correlations of the force field evaluated at different points. 
Such loss of universality is similarly observed in computations of quenched free-energy for disordered systems compared to their annealed counterpart~\cite{crisanti1992spherical} 

The comparison between quenched and annealed counting leads to three main conclusions.
(i) For $\tau \in (0,1]$, a single realization of the disordered system has, with high probability, a total number of equilibria that is exponentially smaller than the ensemble average. The same discrepancy persists when the count is resolved at fixed stability index $\kappa$. 
(ii) The fixed points dominating the annealed count are not the same as those dominating typical realizations: the annealed complexity is maximized by less stable equilibria, namely by fixed points with a larger index $\kappa$, than the quenched complexity, $\kappa_a \geq \kappa_q$. 
(iii) Non-reciprocity suppresses rare-sample bias, while reciprocity enhances it: the discrepancy between annealed and quenched counting grows with $\tau$. Only in the fully non-reciprocal limit, $\tau=0$, do the exponential growth rates of the average and typical number of fixed points coincide and become independent of $\Gamma(q)$. In this limit, the annealed count becomes representative. The bias of the annealed count towards less stable equilibria for $\tau \neq 0$ has a simple origin in this family of models, detailed in the SI. 

Rare events also affect the behavior in the vicinity of the critical lines. The total annealed and quenched complexities both vanish at $\gamma_c=1$, so the annealed count correctly identifies the boundary between the trivial and complex phases. However, it does not generally capture the critical behavior at this boundary: the annealed complexity vanishes quadratically with a fixed prefactor, whereas the quenched complexity has a $\tau$-dependent prefactor and, in the gradient limit $\tau=1$, vanishes cubically. Notice however that the exponent and prefactors are universal with respect to the covariance used in the model.
Thus the transition location is universal, but the typical organization of equilibria around it is not. The discrepancy is more pronounced for the transition between exponentially many and exponentially rare stable equilibria. This transition is not correctly located by the vanishing of the annealed complexity $\Sigma(0)$, which gives \eqref{eq:TauCAnn}. Instead, the quenched transition line $\tau_q(\gamma)$ inherits the explicit dependence on the covariance $\Gamma(q)$ that characterizes typical counting. Although the numerical displacement of the transition line is small in this model, it has a sharp asymptotic meaning: it identifies a narrow region of parameter space, $\tau_a(\gamma)<\tau<\tau_q(\gamma)$, in which stable equilibria are exponentially numerous on average, but their typical number is sub-exponential. Similar annealed-quenched discrepancies in threshold locations have been observed in other high-dimensional disordered systems with random interactions~\cite{ros2023generalized, garnier2021new}.

Beyond its effect on rare-sample bias, non-reciprocity also changes the typical fixed-point population itself. First, increasing reciprocity reduces the total number of equilibria, as the quenched complexity~\eqref{quenched_tot} decreases with $\tau$. Second, it makes the fixed points that dominate the count more stable: their index $\kappa_q$, defined implicitly in \eqref{eq:IndexQ}, decreases with $\tau$; see Fig.~\ref{kappa-star-fig}. This is consistent with the expectation that non-reciprocity tends to destabilize equilibria. 

Our focus on a minimal model has allowed us to analyze explicitly the interplay between non-reciprocity and rare-event effects in ensemble averages. Several directions follow from this work. On the technical side, the quenched complexity framework can be used to access more refined observables, such as the probability that an equilibrium has $k\sim O(N^0)$ unstable directions, as proposed in \cite{ben2021counting}. 
More broadly, a central next step is to connect the fixed-point statistics obtained here with the out-of-equilibrium dynamics of the system. Changes in the organization and stability of equilibria are expected to affect the dynamics, although establishing quantitative connections is a non-trivial task~\cite{fournier2026nonreciprocal}. A natural question is whether the stable fixed points that become exponentially numerous at $\tau_q(\gamma)$, although exponentially subdominant with respect to unstable ones, can qualitatively modify the dynamics in the complex phase, eventually suppressing chaos. The question about dynamics is especially interesting in the region $\tau \in (\tau_a(\gamma), \tau_q(\gamma))$ where stable equilibria are exponentially numerous on average but sub-exponential in typical samples. Addressing this question within this model would be worthwhile, as the phenomenology of this model may extend, at least qualitatively, to ecosystems and neural networks models where related stability transitions in fixed-point statistics are also present~\cite{louis2026fragile}.

\section*{Materials and Methods}
\subsection*{Moments and quenched complexity}
The basis of our computation for the quenched complexity is the expression of the factorial moments for ${\cal N}_{N,k}$ the number of equilibria with $k$ unstable directions. The $n^{\rm th}$ factorial moment ${\cal M}^{n}_{N,k}=\moy{\prod_{p=0}^{n-1}\left({\cal N}_{N,k}-p\right)}$ can be expressed from the Kac-Rice formula as \cite{adler2007random}
\begin{align}
{\cal M}^{n}_{N,k}
={}&\int \prod_{a=1}^n d{\bf x}_a\,\bigg\langle
 \prod_{a=1}^n\delta\!\left({\bf f}({\bf x}_a)\right)\nonumber\\
&\quad\times\prod_{a=1}^n
 |\det({\cal J}({\bf x}_a))|\Theta_k({\cal J}({\bf x}_a))
\bigg\rangle\,,\label{mom-eq}
\end{align}
where the matrix ${\cal J}({\bf x})=\nabla {\bf f}({\bf x})$ is the Jacobian matrix and $\Theta_k(A)$ is an indicator function which is equal to one if the matrix $A$ has exactly $k$ eigenvalues with strictly positive real part and zero otherwise. The factorial moments of ${\cal N}_{N,\rm tot}$ are obtained by removing the indicator functions. An exact expression can be derived for ${\cal M}^1_{N,k}=\moy{{\cal N}_{N,k}}$ in terms of the ratio of a determinant of a deformed matrix from the real Elliptic Gaussian Ensemble and a diagonal matrix containing the relaxation rates \cite{ben2021counting,lacroix2022counting}. This expression can be analysed in any finite dimension $N$ \cite{ben2021counting} (at least for isotropic rates) and the annealed complexity can be extracted from its expression in the high-dimensional limit. The exact expression for $n=1$ stems from the independence of the random field ${\bf f}({\bf x})$ from the Jacobian matrix ${\cal J}({\bf x})$ at coinciding points for stationary random fields. However when considering several (fixed) points, the value of the fields and Jacobian matrix at distinct (fixed) points are strongly correlated, making this computation difficult. In this setting, results are only accessible in the high-dimensional limit where the integrals in \eqref{mom-eq} can be evaluated using a saddle-point method. In the high-dimensional limit and for isotropic\footnote{For anisotropic rates, the computation can be done by considering additionally the matrices with entries $Q_{ab}^1=\sum_i \mu_i x_{ai}x_{bi}$ and $Q_{ab}^2=\sum_i \mu_i^2 x_{ai}x_{bi}$.} rates $\mu_i=\mu$, one expects that the integrand in \eqref{mom-eq} depends, at leading exponential order, only on the overlap matrix $Q_{ab}={\bf x}_a\cdot{\bf x}_b$. The integral in \eqref{mom-eq} is expressed as
\begin{align}
{\cal M}^{n}_{N,k=N\kappa}&=\int_{Q>0} dQ\,r_n(Q)\,e^{N\left[P_n(Q)+V_n(Q)+J_n(Q,\kappa)\right]}\,,
\end{align}
where the integral is over positive definite matrices $Q$ and the probability $P_n(Q)$, volume $V_n(Q)$ and Jacobian $J_n(Q,\kappa)$ terms read respectively
{\small
\begin{align}
    P_n&=\lim_{N\to \infty}\frac{1}{N}\ln\moy{\prod_{a=1}^n\delta\left({\bf f}({\bf x}_a)\right)}\,,\\
    V_n&=\lim_{N\to \infty}\frac{1}{N}\ln\int \prod_{a=1}^n d{\bf x}_a\prod_{a,b}\delta\left(Q_{ab}-{\bf x}_a\cdot{\bf x}_b\right)\,,\\
    J_n&=\lim_{N\to \infty}\frac{1}{N}\ln\moy{\prod_{a=1}^n |\det({\cal J}({\bf x}_a))|\Theta_{N\kappa}({\cal J}({\bf x}_a))\Big|{\bf f}({\bf x}_a)={\bf 0}}\,,
\end{align}
}
where in the last expression the average product of Jacobians at points ${\bf x}_a$'s is conditioned on vanishing fields at each point. 
Using the saddle-point method, one obtains 
\be
\lim_{N\to \infty}\frac{1}{N}\ln{\cal M}^{n}_{N,N\kappa}=\max_{Q>0}\left[P_n(Q)+V_n(Q)+J_n(Q,\kappa)\right]\,.
\ee
Each of the three terms can be computed for any value of $n$ and without specific assumption on the structure of $Q$ in the special case $\tau=0$ of a divergence-free field, yielding the result in \eqref{div-free}. For any $\tau>0$ instead, we have performed the computations of the probability and Jacobian terms assuming that the overlap matrix is replica-symmetric $Q_{ab}=(q_d-q)\delta_{ab}+q$. In the limit $n\to 0$, one obtains
\be
\hat \Xi(\kappa)=\lim_{N\to \infty}\frac{\moy{\psi({\cal N}_{N,N\kappa}+1)}}{N}=\lim_{n\to 0}\lim_{N\to \infty}\frac{1}{nN}\ln{\cal M}^{n}_{N\kappa}\,,
\ee
where $\psi(x)$ is the digamma function. This expression is our proxy to obtain the quenched complexity. Indeed using the asymptotic behaviour of that function $\psi(x)\approx \ln x$ as $x\gg 1$, one sees that the expression above coincides with the quenched complexity, i.e. $\hat \Xi(\kappa)=\Xi(\kappa)\geq 0$. In addition to the asymptotic expression of the average $\moy{\psi({\cal N}_{N,N\kappa}+1)}$, we also consider in the SI its variance showing that it grows linearly with $N$ as well and computing exactly its rescaled expression ${\cal V}(\kappa)=\lim_{N\to \infty}{\rm Var}\left(\psi({\cal N}_{N,N\kappa}+1)\right)/N$. Using the asymptotic behaviour of the digamma function, we can conclude that the fluctuations of $\ln({\cal N}_{N,N\kappa}+1)$ are of order $O(\sqrt{N})$ and their rescaled variance is ${\cal V}(\kappa)$.

\subsection*{Count at fixed extensive index} 
This family of models allows us to calculate the complexity of equilibria at fixed extensive index $k=N \kappa$. The crucial observation is that each Jacobian ${\cal J}({\bf x}_a)$ in \eqref{mom-eq}, conditioned on ${\bf x}_b$ being a fixed point for all $b \in \{1, \cdots, n\}$, is a Gaussian random matrix. For fixed ${\bf x}_a$, it has the same marginal law as the sum
\be
{\cal J}({\bf x}_a)\underset{\rm law}{\equiv}-D^\mu+X^a+\xi_a\mathbb{I}+W^a \;.\label{eq:prob_jac_MM}
\ee
Here $D^\mu$ is a deterministic matrix with components $D^\mu_{ij}=\mu_i \delta_{ij}$. Each other term in this decomposition has entries with Gaussian statistics: the Jacobians prior to conditioning are Gaussian, and conditioning on the linear constraints $F_i({\bf x}_b)=\mu_i x_{bi}$ preserves Gaussianity. The $X^a$'s are matrices drawn from the real Elliptic Gaussian Ensemble (rEGE) with $\moy{X_{ij}^a}=0$, $\moy{X_{ij}^a X^{a}_{kl}}=c\left(\delta_{ik}\delta_{jl}+\tau\delta_{il}\delta_{jk}\right)$ and $c=4(\Gamma_V''(0)+\Gamma_A''(0))$, $\tau=4\Gamma_V''(0)/c$. This part of the Jacobian is unaffected by the conditioning. The $\xi_a$'s are Gaussian scalars, independent of the $X^a$'s, with mean of order $N^{{1}/{2}}$ (comparable to $\mu$ and to the width of the spectrum of $X^a$), and covariance of order one. Finally, $W^a$ is a random matrix of finite rank, correlated with both the matrices $X^a$'s and the random variables $\xi_a$'s. It is absent for a single, unconditioned Jacobian, and it arises from conditioning on the other fixed points. 

The diagonal shift in \eqref{eq:prob_jac_MM} is present also prior to conditioning. In fact, the structure $X^a+\xi_a\mathbb{I}$ directly reflects the symmetry of the force-field covariance: because it is translationally invariant, the covariance of the Hessian of $V$ is the fourth derivative of a single function, hence totally symmetric in its indices; isotropy then forces it to be
$4\Gamma_V''(0)(\delta_{ik}\delta_{jl}+\delta_{il}\delta_{jk}
+\delta_{ij}\delta_{kl})$. The divergence-free part contributes
$4\Gamma_A''(0)\,\delta_{ik}\delta_{jl}$. In total, $\langle\mathcal J_{ij}\mathcal J_{kl}\rangle_c
= c\,(\delta_{ik}\delta_{jl}+\tau\delta_{il}\delta_{jk})
+ c\,\tau\,\delta_{ij}\delta_{kl}.$
The first term gives $X^a$; the second gives the shift $\xi_a$, which comes entirely from the gradient part and vanishes for $\tau=0$. When conditioning on  $F_i({\bf x}_b)=\mu_i x_{bi}$, the statistics of the $\xi_a$'s is modified, as discussed in detail in the SI.

Fixing an index $k_a$ for a fixed point ${\bf x}_a$ corresponds to requiring that the matrix \eqref{eq:prob_jac_MM} has $k$ eigenvalues with positive real part. In particular, given $\rho_{a,N}(z)$ the eigenvalue distribution of ${\cal J}({\bf x}_a)-\xi_a \mathbb{I}$,  then
\begin{align}
k_a&=N\int_{\Re(z)+\xi_a>0} d^2 z\,\rho_{a,N}(z)\;.
\end{align}
In the scaling 
$k=N\kappa$ with $\kappa>0$, changing $\kappa$ is achieved by either shifting $\xi_a$ by an amount of order $ N^{{1}/{2}}$, or by deforming $\rho_{a,N}(z)$. The first large deviation has a probability that is only exponentially small in $N$, while the second one decays much faster. Therefore, the counting of fixed points at fixed extensive index is achieved simply by conditioning on the value of the Gaussian variables $\xi_a$. Notice that to leading order $\xi_a= N^{-1} \Tr[{\cal J}({\bf x}_a)+D^\mu]= N^{-1} \sum_{i=1}^N \partial_i F_i({\bf x}_a)$: therefore, the stability of equilibria is controlled by a single collective quantity, the local divergence of the random force field. Given the decomposition \eqref{eq:ForceDec}, only the potential $V$ contributes to the divergence: $\xi_a= -N^{-1} \Delta V({\bf x}_a) $. Thus, for $\tau=0$ (divergence-free field), the shift disappears and the complexity is positive only for $\kappa=\kappa_q$. 
In the SI we argue that the divergence $\xi_a= -N^{-1} \Delta V({\bf x}_a)$ is largely shared by all the equilibria of a given realization, and it is larger for realizations rich in equilibria. These are the realizations that dominate the annealed counting, that is thus biased towards less stable equilibria (larger shifts $\xi_a$). This mechanism is absent for divergence-free forces, where indeed quenched and annealed counts coincide. 

\subsection*{Stability of the replica symmetric assumption}
The expressions for the quenched complexities, \eqref{quenched_tot} and \eqref{quenched_kappa}, are obtained under the assumption of replica symmetry, encoded in the structure of the matrix $Q$. The stability of the replica symmetric assumption with respect to perturbations breaking continuously the symmetric structure of $Q$ can nevertheless be checked: we find that the stability holds whenever the quantity 
{\small
\begin{equation}\label{eq:Replicon}
\begin{split}
\Lambda_r&=\frac{2(1-\gamma v(\gamma))^2}{{\cal Q}(\gamma)^2}\left[1+\frac{\gamma A^2}{\tau(1-v(\gamma))^2}\left(v''(\gamma)-\frac{2v(\gamma)v'(\gamma)}{1-\gamma v(\gamma)}\right)\right]\\
&-\frac{4(1-\gamma v(\gamma))^2}{{\cal Q}(\gamma)^2(1+\tau\gamma v(\gamma))}\left(1-\frac{A\gamma}{1-v(\gamma)}v'(\gamma)\right)^2,\\
A&=1+\tau\gamma-\sqrt{\gamma}(1+\tau)z(\kappa)\;
\end{split}
\end{equation}
}
is negative. In the language of replica theory, $\Lambda_r$ is the so called replicon eigenvalue. When evaluating this expression for the total complexity, one should consider $z \to z(\kappa_q)$, and thus $A \to A_*$ with
\be
A_*=\frac{\tau(\gamma-1)(1-v(\gamma))}{1+\tau v(\gamma)}\;,
\ee
while one should take $z(\kappa=0)=1$ when evaluating $\Lambda_r$ for stable equilibria. For the correlation functions and in the regime of parameters used in the main text, we have checked explicitly that $\Lambda_r<0$ (see SI for details).

\section*{Acknowledgments}
The authors thank Yan V. Fyodorov for useful discussions. VR acknowledges funding by the French government under the France 2030 program (PhOM - Graduate
School of Physics) with reference ANR-11-IDEX-0003. BLACT and VR acknowledge funding from PhOM-V 2024 project NAMHiDiCS and BLACT is grateful for the hospitality of LPTMS where this project was initiated.
PLD acknowledges support of the ANR Grant No. ANR23-CE30-0020-01 EDIPS. The authors acknowledge support from CNRS IRP Hidings, and are grateful for the hospitality of the
program ``The Lush World of Random Matrices" at the ZiF center of Bielefeld
University, where this work has been finalized.

%% file: supplement.tex
\section{Quenched complexity}\label{sec:RKR}
Below, we provide a detailed derivation of our expression for the $n^{\rm th}$ factorial moment of the number of equilibria with $k$ unstable directions. The factorial moments are expressed as \cite{adler2007random}
\begin{align}
{\cal M}^{n}_{k}=&\moy{\prod_{p=0}^{n-1}\left({\cal N}_{N,k}-p\right)}=\int \prod_{a=1}^n d{\bf x}_a\,\moy{\prod_{a=1}^n\delta\left({\bf f}({\bf x}_a)\right)\prod_{a=1}^n |\det({\cal J}({\bf x}_a))|\Theta_k({\cal J}({\bf x}_a))}\;,\label{mom_n}\\
=&\int \prod_{a=1}^n d{\bf x}_a\,\moy{\prod_{a=1}^n\delta\left({\bf f}({\bf x}_a)\right)}\moy{\prod_{a=1}^n |\det({\cal J}({\bf x}_a))|\Theta_k({\cal J}({\bf x}_a))\Bigg|\forall a=1,\cdots,n\;,\;\;{\bf f}({\bf x}_a)={\bf 0}}\nn
\end{align}
where 
\be\label{eq:DefJ}
f_i({\bf x})=-\mu_i x_i-\partial_{x_i}V({\bf x})+\frac{1}{\sqrt{N}}\sum_{l=1}^N \partial_{x_l}A_{il}({\bf x})\;\;{\rm and}\;\;{\cal J}_{ij}({\bf x})=\partial_{x_j}f_i({\bf x})=-\mu_{i}\delta_{ij}+J_{ij}({\bf x})=(-D^{\mu}+J({\bf x}))_{ij}\;,
\ee
with $D_{ij}^{\mu}=\mu_{i}\delta_{ij}$ and $J_{ij}({\bf x})=\partial_j F_i({\bf x})$. The expectation $\moy{\cdots}$ above is taken with respect to the Gaussian potential $V({\bf x})$ and the asymmetric Gaussian matrix field $A({\bf x})$. The latter are Gaussian, independent of each other, have zero average and covariance
\begin{align}
    \moy{V({\bf x}_a)V({\bf x}_b)}&=\Gamma_V\left(|{\bf x}_a-{\bf x}_b|^2\right)\,,\\
    \moy{A_{ij}({\bf x}_a)A_{kl}({\bf x}_b)}&=\left(\delta_{ik}\delta_{jl}-\delta_{il}\delta_{jk}\right)\Gamma_A\left(|{\bf x}_a-{\bf x}_b|^2\right)\,.
\end{align}
For convenience, we also introduce the notation ${\bf m}_a$ such that
\be
m_{ai}=\mu_i x_{ai}\;,\;\;a=1,\cdots,n\;,\;\;i=1,\cdots,N\;.
\ee
The expression in \eqref{mom_n} can be reformulated as an integral over the overlap matrix 
\be
\mathbb{Q}=\begin{pmatrix}
    Q^0&Q^1\\
    Q^1&Q^2
\end{pmatrix}\;,\;\;Q^{\eta}_{ab}=\sum_{i=1}^N \mu_i^{\eta}\,x_{ai}x_{bi}\;,\;\;\eta=0,1,2\;,\;\;a,b=1,\cdots,n\;.
\ee
Notice that when $\mu_i=\mu$, then $Q^\eta=\mu^\eta Q^0$ and therefore only $Q^0$ has to be integrated over.
With the covariance convention above, a nontrivial competition between the disorder and relaxation rates requires $\mu_i=O(N^{1/2})$.
In order to only rely on quantities of order $O(1)$ in the computation that follows, and take properly the high-dimensional limit $N\to \infty$, we introduce the rescaled quantities
\begin{align}
    &\hat {\bf x}=\sqrt{N}{\bf x}\;,\;\;\hat\mu_i=\frac{\mu_i}{\sqrt{N}}\;,\;\;\hat V(\hat {\bf x})=\sqrt{N}V\left(\frac{\hat {\bf x}}{\sqrt{N}}\right)\;,\;\;\hat A_{ij}(\hat {\bf x})=\sqrt{N}A_{ij}\left(\frac{\hat {\bf x}}{\sqrt{N}}\right)\;,\label{c1}\\
    &\hat f_i(\hat {\bf x})=f_i\left(\frac{\hat {\bf x}}{\sqrt{N}}\right)=-\hat \mu_i \hat x_i-\partial_{\hat x_i}\hat V(\hat {\bf x})+\frac{1}{\sqrt{N}}\sum_{l=1}^N \partial_{\hat x_l}\hat A_{il}(\hat {\bf x})\;,\;\;\hat {\cal J}_{ij}(\hat {\bf x})=\partial_{\hat x_j}\hat f_i(\hat {\bf x})=\frac{1}{\sqrt{N}}\partial_{x_j}f_i({\bf x})=\frac{{\cal J}_{ij}({\bf x})}{\sqrt{N}}\;,\label{c2}\\
    &\hat m_{ai}=\hat \mu_i \hat x_{ai}=m_{ai}\;,\;\;\hat Q^{\eta}_{ab}=\frac{1}{N}\sum_{i=1}^N \hat \mu_i^{\eta}\,\hat x_{ai}\hat x_{bi}=N^{-\eta/2}Q_{ab}^{\eta}\;.\label{c3}
\end{align}
In particular, one has that
\be
d{\bf x}_a\delta\left({\bf f}({\bf x}_a)\right)|\det({\cal J}({\bf x}_a))|=d\hat {\bf x}_a\delta\left(\hat {\bf f}(\hat {\bf x}_a)\right)|\det(\hat {\cal J}(\hat{\bf x}_a))|\;,
\ee
allowing to work with the rescaled variables to compute the factorial moments in the expression above. Finally, with this convention, we recover the usual scaling of the correlation functions used in the disordered systems literature
\begin{align}
    \moy{\hat V(\hat {\bf x}_a)\hat V(\hat {\bf x}_b)}&=N\Gamma_V\left(\frac{|\hat{\bf x}_a-\hat{\bf x}_b|^2}{N}\right)\,,\\
    \moy{\hat A_{ij}(\hat{\bf x}_a)\hat A_{kl}(\hat{\bf x}_b)}&=\left(\delta_{ik}\delta_{jl}-\delta_{il}\delta_{jk}\right)N\Gamma_A\left(\frac{|\hat{\bf x}_a-\hat{\bf x}_b|^2}{N}\right)\,.
\end{align}
With this rescaling, one can introduce the empirical density of relaxation rates  
\be\label{eq:DenMU}
\nu_N(\hat \mu)=\frac{1}{N}\sum_{i=1}^N \delta\left(\hat \mu-\hat\mu_i\right)\;,
\ee
which we assume converges to a well-defined measure $\nu(\hat \mu)$ as $N\to \infty$. We work within these rescaled notations in the computations below and for convenience of the notations, we now drop all the hat notations. The original scaling will be restored in the final expressions.

In the new convention with rescaled variables, we can express
\begin{align}
{\cal M}^{n}_{k}=\moy{\prod_{p=0}^{n-1}\left({\cal N}_{N,k}-p\right)}=\int d\mathbb{Q}\,{\cal V}_{N,n}(\mathbb{Q}){\cal P}_{N,n}(\mathbb{Q}){\cal E}_{N,n,k}(\mathbb{Q})\;,\;\;{\rm with}\;\;d\mathbb{Q}=dQ^0dQ^1dQ^2\;\;{\rm and}\;\;dQ^{\eta}=\prod_{a\leq b}dQ_{ab}^{\eta}\;.\label{Nkn}
\end{align}
The expression above is obtained by introducing the volume element 
\be
{\cal V}_{N,n}(\mathbb{Q})=\int \prod_{a=1}^n d{\bf x}_a\,\prod_{a \leq b=1}^n\prod_{\eta=0}^2\delta\left(\frac{1}{N}\sum_{i=1}^N \mu_i^{\eta}\,x_{ai}x_{bi}-Q_{ab}^{\eta}\right)\;.
\ee
The second term corresponds to the joint distribution of the disordered fields 
\begin{align}
 {\cal P}_{N,n}(\mathbb{Q})=&\moy{\prod_{a=1}^n \prod_{i=1}^N \delta\left(f_i({\bf x}_a)\right)}\;,
\end{align}
which is similarly only expected to depend on the overlaps.
Finally, the third term is the conditional expectation of product of determinants which also depends on the overlaps and reads
\be\label{eq:CondHess}
{\cal E}_{N,n,k}(\mathbb{Q})=\moy{\prod_{a=1}^n |\det({\cal J}({\bf x}_a))|\Theta_k({\cal J}({\bf x}_a))\Bigg|\forall a=1,\cdots,n\;,\;\;{\bf f}({\bf x}_a)={\bf 0}}\;.
\ee

In the following, we provide the general expression for these three different terms whenever possible before specialising to the replica-symmetric (henceforth, RS) ansatz which corresponds to
\be
Q_{ab}^{\eta}=q_{\eta}-\Delta_{\eta}(1-\delta_{ab})\;,\;\;\eta=0,1,2\;,\;\;a,b=1,\cdots,n\;.
\ee

\subsection{Volume element}

Let us first obtain an explicit expression for the volume element ${\cal V}_{N,n}(\mathbb{Q})$ in the replica-symmetric ansatz. First we introduce an integral representation of the delta functions to obtain
\begin{align}
{\cal V}_{N,n}(\mathbb{Q})
&=C_{N,n}\int \prod_{a=1}^n d{\bf x}_a\,\int \prod_{a,b=1}^n\prod_{\eta=0}^2\frac{d\Lambda_{ab}^\eta}{2\pi}\,e^{-\frac{iN}{2}\left[\frac{1}{N}\sum_{j}\sum_{a,b}x_{aj}\sum_{\eta=0}^2\mu_j^{\eta}\Lambda_{ab}^\eta x_{bj}-{\rm Tr}\left(\sum_{\eta=0}^2\Lambda^{\eta}Q^{\eta}\right)\right]}\\
&=C_{N,n}\int \prod_{a,b=1}^n\prod_{\eta=0}^2\frac{d\Lambda_{ab}^\eta}{2\pi}\exp\left(\frac{N}{2}\left[{\rm Tr}\left(\sum_{\eta=0}^2 i\Lambda^{\eta}Q^{\eta}\right)-\frac{1}{N}\sum_{j=1}^N\ln \det\left(\sum_{\eta=0}^2\mu_j^{\eta}i\Lambda^{\eta}\right)+n\ln(2\pi)\right]\right)\nn\\
&=\tilde C_{N,n}\int\prod_{a,b=1}^n\prod_{\eta=0}^2\frac{dL_{ab}^\eta}{2\pi}\exp\left(\frac{N}{2}\left[{\rm Tr}\left(\sum_{\eta=0}^2 L^{\eta}Q^{\eta}\right)-\frac{1}{N}\sum_{j=1}^N\ln \det\left(\sum_{\eta=0}^2\mu_j^{\eta}L^{\eta}\right)+n\ln(2\pi)\right]\right)\;,\nn
\end{align}
with $\lim_{N\to \infty}\frac{1}{N}\ln \tilde C_{N,n}=0$. This last expression can be evaluated using the saddle-point method. The saddle-point equations read
\begin{align}
    Q_{ab}^{\eta}&=\frac{1}{N}\sum_{j=1}^N\mu_j^{\eta} \left(\sum_{\omega=0}^2\mu_j^{\omega} L^\omega\right)^{-1}_{ab}\;,\;\;\eta=0,1,2\;,\;\;a,b=1,\cdots,n\;.
\end{align}
Using these identities, one can obtain
\be
{\rm Tr}\left(\sum_{\eta=0}^2 L^{\eta}Q^{\eta}\right)={\rm Tr}\left(\mathbb{I}\right)=n\;.
\ee

Introducing the RS ansatz,
\be
L_{ab}^\eta=l_{\eta}-\lambda_\eta(1-\delta_{ab})\;,\;\;\eta=0,1,2\;,\;\;a,b=1,\;\cdots,n\;,
\ee
the saddle-point equations take the form 
\begin{align}
    q_{\eta}-\Delta_{\eta}(1-\delta_{ab})&=\int d\mu\,\frac{\mu^{\eta}\nu_N(\mu)}{\sum_{\omega=0}^2\mu^{\omega}\lambda_{\omega}}\,\left[1-\frac{\sum_{\omega=0}^2\mu^{\omega}(l_{\omega}-\lambda_{\omega})}{\sum_{\omega=0}^2\mu^{\omega}[\lambda_{\omega}+n(l_{\omega}-\lambda_{\omega})]}-(1-\delta_{ab})\right]\;,\label{q_lambda}
\end{align}
where we have used the empirical distribution  $\nu_N(\mu)$ defined in \eqref{eq:DenMU}. The leading asymptotic behaviour of the volume term in the RS ansatz as 
\begin{align}
    \lim_{N\to \infty}&\frac{1}{N}\ln{\cal V}_{N,n}(\mathbb{Q})=\frac{n}{2}\left(1+\ln(2\pi)-\int d\mu\, \nu(\mu)\,\ln\left[\sum_{\eta=0}^2\mu^{\eta}\lambda_{\eta}\right]\right)-\frac{1}{2}\int d\mu\,\nu(\mu)\,\ln\left[1+n\left(\frac{\sum_{\eta=0}^2\mu^{\eta}l_{\eta}}{\sum_{\eta=0}^2\mu^{\eta}\lambda_{\eta}}-1\right)\right]\;.\label{vol_term}
\end{align}
In the large $N$ limit, the relation between the overlap parameters and the Lagrange parameters is given by \eqref{q_lambda} where the empirical distribution $\nu_N(\mu)$ is replaced by the limiting distribution $\nu(\mu)=\lim_{N\to \infty}\nu_N(\mu)$. For isotropic relaxation rates $\nu(\mu')=\delta(\mu-\mu')$, such that $Q^{\eta}=\mu^{\eta}Q$ and $L=\sum_{\eta}\mu^{\eta}L^{\eta}$, the saddle-point equation reads $Q=L^{-1}$, such that
\begin{align}
    \lambda=\frac{1}{\Delta}\;,\;\;l=\frac{1}{n}\left[\frac{n-1}{\Delta}+\frac{1}{\Delta+n(q-\Delta)}\right]\;,\;\;\lim_{N\to \infty}&\frac{1}{N}\ln{\cal V}_{N,n}=\frac{n}{2}\left(1+\ln(2\pi\Delta)\right)+\frac{1}{2}\ln\left[1+n\left(\frac{q}{\Delta}-1\right)\right]\;,
\end{align}
with $\Delta_0=\Delta_1/\mu=\Delta_2/\mu^2=\Delta$ and $q_0=q_1/\mu=q_2/\mu^2=q$. 

\subsection{Distribution of fields} We now evaluate the Gaussian probability ${\cal P}_{N,n}(\mathbb{Q})$ of the non-gradient field, within the RS ansatz.
We use that the fields have Gaussian statistics with zero average $\moy{{\bf F}({\bf x}_a)}=0$ and covariance
\begin{align}
\Sigma_{ij}^{ab}=&\moy{F_i({\bf x}_a)F_j({\bf x}_b)}=\moy{\partial_{x_{ai}}V({\bf x}_a)\partial_{x_{bj}}V({\bf x}_b)}+\frac{1}{N}\sum_{k,l}\moy{\partial_{x_{ak}}A_{ik}({\bf x}_a)\partial_{x_{bl}}A_{jl}({\bf x}_b)}\label{sigma_matrix}\\
=&N\partial_{x_{ai},x_{bj}}^2 \Gamma_V\left(\frac{|{\bf x}_a-{\bf x}_b|^2}{N}\right)+\sum_{k,l}(\delta_{ij}\delta_{kl}-\delta_{il}\delta_{jk})\partial_{x_{ak},x_{bl}}^2 \Gamma_A\left(\frac{|{\bf x}_a-{\bf x}_b|^2}{N}\right)\nn\\
=&-\delta_{ij}D\left(d_{ab}^{2}\right)-\frac{d_i^{ab}d_j^{ab}}{N}R\left(d_{ab}^{2}\right),\label{sigma_matrixb}
\end{align}
with 
\be\label{CrucialTerms}
D(q)=2(\Gamma_{V}'\left( q\right)+\Gamma_{A}'\left(q\right))-\frac{2}{N}(\Gamma_{A}'\left(q\right)-2q\Gamma_{A}''\left(q\right))\;, \quad\;\;R( q)=4\Gamma_{V}''\left(q\right)-\frac{4}{N}\Gamma_{A}''\left(q\right)\\\;\;,
\ee
where for convenience of the notations we have introduced
\be
d_i^{ab}=x_{ai}-x_{bi}\;,\quad\;\;d_{ab}^{2}=\frac{1}{N}\sum_{k=1}^N \left(d_k^{ab}\right)^2\;.\label{dab}
\ee
The joint distribution of fields then reads
\begin{align}\label{eq:QF}
 {\cal P}_{N,n}(\mathbb{Q})=&\moy{\prod_{a=1}^n \prod_{i=1}^N \delta\left(f_i({\bf x}_a)\right)}=\frac{1}{(2\pi)^{nN/2}\sqrt{\det \Sigma}}\,\exp\left(-\frac{1}{2}\sum_{i,j=1}^N \sum_{a,b=1}^n m_{ai} \left(\Sigma^{-1}\right)^{ab}_{ij} m_{bj}\right)\;,
\end{align}
where we remind that $m_{ai}=\mu_i x_{ai}$. Note that the last term appearing in \eqref{sigma_matrixb} is a finite-rank perturbation of the first, full-rank term, and will lead to a sub-leading contribution when evaluating $\ln \det \Sigma$. However, it cannot be neglected in the calculation of the quadratic form in \eqref{eq:QF} because the vectors ${\bf m}_a$'s and ${\bf x}_b$'s have non-trivial overlaps. The expression above is valid for any number of replica-symmetry breaking. The inverse of the covariance function $\Sigma^{-1}$, appearing in the Gaussian term is however hard to obtain explicitly in its general form.

We now show how the quadratic term can be expressed within the replica-symmetric ansatz, where
\be
\frac{{\bf x}_a\cdot{\bf x}_b}{N}=Q_{ab}^0=q_0-\Delta_0(1-\delta_{ab})\;,\quad \;\;d_{ab}^2=\frac{({\bf x}_a-{\bf x}_b)^2}{N}=2\Delta_0(1-\delta_{ab})\;.
\ee
We make use of the fact that the four-dimensional sub-space $\{{\bf m}_a,\sum_{c\neq a}{\bf m}_c,{\bf x}_a,\sum_{c\neq a}{\bf x}_c\}$, spanned by vectors satisfying the following identities within the replica-symmetric scheme
\begin{align}
    \frac{{\bf m}_a\cdot{\bf m}_b}{N}&=q_2-(1-\delta_{ab})\Delta_2\;,\;\;\frac{{\bf m}_a\cdot\sum_{c\neq b}{\bf m}_c}{N}=(n-1)(q_2-\Delta_2)+(1-\delta_{ab})\Delta_2\;,\label{dotprod1}\\
    \frac{{\bf m}_a\cdot{\bf x}_b}{N}&=q_1-(1-\delta_{ab})\Delta_1\;,\;\;\frac{{\bf m}_a\cdot\sum_{c\neq b}{\bf x}_c}{N}=(n-1)(q_1-\Delta_1)+(1-\delta_{ab})\Delta_1\;,\label{dotprod2}
\end{align}
is stable under multiplication by the covariance matrix, i.e.
\begin{align}
    \sum_{a,i}m_{ai}\Sigma_{ij}^{ab}=&-D(0)m_{bj}-D(2\Delta_0)\sum_{a\neq b}m_{aj}+\Delta_1 R(2\Delta_0)\left((n-1)x_{bj}-\sum_{a\neq b}x_{aj}\right)\;,\\
    \sum_{a,i}\sum_{c\neq a}m_{ci}\Sigma_{ij}^{ab}=&-D(2\Delta_0)(n-1)m_{bj}-(D(0)+D(2\Delta_0)(n-2))\sum_{c\neq b}m_{cj}-\Delta_1 R(2\Delta_0)\left((n-1) x_{bj}-\sum_{c\neq b} x_{cj}\right)\;,\\
    \sum_{a,i}x_{ai}\Sigma_{ij}^{ab}=&\left(-D(0)+(n-1)\Delta_0 R(2\Delta_0)\right) x_{bj}-\left(D(2\Delta_0)+\Delta_0 R(2\Delta_0)\right)\sum_{a\neq b} x_{aj}\;,\\
    \sum_{a,i}\sum_{c\neq a} x_{ci}\Sigma_{ij}^{ab}=&-(n-1)\left(D(2\Delta_0)+\Delta_0 R(2\Delta_0) \right)x_{bj}-(D(0)+D(2\Delta_0)(n-2)-\Delta_0 R(2\Delta_0))\sum_{c\neq b} x_{cj}\;.
\end{align}

The covariance matrix can thus be expressed explicitly in this sub-space. Note that, as shown by the identities in Eqs. (\ref{dotprod1}-\ref{dotprod2}), the vectors are not orthonormal. Nevertheless, the action of the inverse of the covariance matrix on these vectors can be computed explicitly, leading to the following expression 
\begin{align}
\sum_{a,i}m_{ai}\left(\Sigma^{-1}\right)_{ij}^{ab}=&\frac{1}{(D(0)-D(2\Delta_0))(D(0)+(n-1)D(2\Delta_0))}\left[-\left(D(0)+(n-2)D(2\Delta_0)\right)m_{bj}+D(2\Delta_0)\sum_{c\neq b}m_{cj}\right]\label{inversesig_m}\\
&+\frac{1}{(D(0)-D(2\Delta_0))(D(0)-D(2\Delta_0)-n \Delta_0 R(2\Delta_0))}\left[-(n-1)\Delta_1 R(2\Delta_0) x_{bj}+\Delta_1 R(2\Delta_0)\sum_{c\neq b}x_{cj}\right]\;.\nn
\end{align}
Finally, within the replica-symmetric ansatz,  one obtains the following explicit expression for the joint distribution of fields
{\small
\begin{align}
    &\lim_{N\to \infty}\frac{1}{N}\ln{\cal P}_{N,n}=-\frac{n}{2}\ln(2\pi)-\frac{1}{2}\left[n\ln\left(D(2\Delta_0)-D(0)\right)+\ln\left(1-\frac{nD(2\Delta_0)}{D(2\Delta_0)-D(0)}\right)\right]\label{Gau_term}\\
    &+\frac{n}{2}\left(\frac{q_2}{D(0)+(n-1)D(2\Delta_0)}+\frac{(n-1)\Delta_2 D(2\Delta_0)}{(D(0)+(n-1)D(2\Delta_0))(D(0)-D(2\Delta_0))}+\frac{(n-1)\Delta_1^2 R(2\Delta_0) }{(D(0)-D(2\Delta_0)-n\Delta_0 R(2\Delta_0))(D(0)-D(2\Delta_0))}\right)\;.\nn
\end{align}
}
For isotropic rates, denoting $\Delta_0=\Delta_1/\mu=\Delta_2/\mu^2=\Delta$, one obtains 
{\small
\begin{align}
    &\lim_{N\to \infty}\frac{1}{N}\ln{\cal P}_{N,n}=-\frac{n}{2}\ln(2\pi)-\frac{1}{2}\left[n\ln\left(D(2\Delta)-D(0)\right)+\ln\left(1-\frac{nD(2\Delta)}{D(2\Delta)-D(0)}\right)\right]\\
    &+\frac{n\mu^2}{2}\left(\frac{q}{D(0)+(n-1)D(2\Delta)}+\frac{(n-1)\Delta D(2\Delta) }{(D(0)+(n-1)D(2\Delta))(D(0)-D(2\Delta))}+\frac{(n-1)\Delta^2 R(2\Delta)}{(D(0)-D(2\Delta)-n\Delta R(2\Delta))(D(0)-D(2\Delta))}\right)\;.\nn
\end{align}
}
Note that in this limit, the $1/N$ corrections to $D(q)$ and $R(q)$, see \eqref{CrucialTerms}, are dropped. 

\subsection{Jacobian term} Finally, we consider the Jacobian term ${\cal E}_{N,n,k}(\mathbb{Q})$ in the replica-symmetric ansatz. To address this term, we first derive the statistics of the matrices $J({\bf x}_a)$ defined in \eqref{eq:DefJ} conditioned on all ${\bf x}_a$'s being fixed points. This will allow us to treat the Jacobian term ${\cal E}_{N,n,k}(\mathbb{Q})$, within the replica-symmetric ansatz. The un-conditioned covariance between the field and the Jacobian entries reads
\begin{align}
\moy{J_{ij}({\bf x}_a)F_k({\bf x}_b)}=
&-\frac{\delta_{ik}d_j^{ab}}{N}2D'\left(d_{ab}^{2}\right)-\frac{\delta_{ij}d_k^{ab}+\delta_{jk}d_i^{ab}}{N}R\left(d_{ab}^{2}\right)-\frac{d_i^{ab}d_j^{ab}d_k^{ab}}{N^2}2R'\left(d_{ab}^{2}\right)\;,
\end{align}
where $d_{i}^{ab}$ and $d_{ab}^{2}$ are defined in \eqref{dab}. Similarly, the un-conditioned covariance of Jacobian entries reads
\begin{align}
\moy{J_{ij}({\bf x}_a)J_{kl}({\bf x}_b)}=&
\frac{\delta_{ij}\delta_{kl}+\delta_{jk}\delta_{il}}{N}R\left(d_{ab}^{2}\right)+\frac{\delta_{ik}\delta_{jl}}{N} 2D'\left(d_{ab}^{2}\right)+\frac{\delta_{ik}d_j^{ab}d_l^{ab}}{N^2}4D''\left(d_{ab}^{2}\right)+ \frac{d_i^{ab}d_j^{ab}d_k^{ab}d_l^{ab}}{N^3}4R''\left(d_{ab}^{2}\right)\nn\\
&+\frac{\delta_{ij}d_k^{ab}d_l^{ab}+\delta_{jk}d_i^{ab}d_l^{ab}+\delta_{il}d_k^{ab}d_j^{ab}+\delta_{jl}d_i^{ab}d_k^{ab}+\delta_{lk}d_i^{ab}d_j^{ab}}{N^2}2R'\left(d_{ab}^{2}\right)\;.
\end{align}
Note that only the first three terms of this covariance are full-rank. In particular, the second and the third terms correspond to the statistics of random matrices of the Gaussian Elliptic type, correlated with each other for different $a$; the first one is the covariance that one would get for diagonal random matrices with uniform Gaussian entries, again correlated to each other for different $a$. The remaining terms correspond to finite-rank perturbations: in particular, these terms give a contribution only to the covariances of matrix elements of $J({\bf x}_a)$ along vectors that are linear combinations of the ${\bf x}_a$: the number of such matrix elements is sub-leading with respect to the total number of entries of the matrices.

The conditional distribution remains Gaussian with the conditional expectation and covariance reading
\begin{align}
&\moy{J_{ij}({\bf x}_a)|\forall b\in\{1,\cdots,n\},{\bf F}({\bf x}_b)={\bf m}_b}=\sum_{b,c}\sum_{k,l}\moy{F_k({\bf x}_b)J_{ij}({\bf x}_a)}\left(\Sigma^{-1}\right)^{bc}_{kl}m_{cl}\;,\label{cond_avg}\\
    &{\rm Cov}\left[J_{ij}({\bf x}_a),J_{kl}({\bf x}_b)|\forall c\in\{1,\cdots,n\},{\bf F}({\bf x}_c)={\bf m}_{c}\right]=\moy{J_{ij}({\bf x}_a)J_{kl}({\bf x}_b)}-\sum_{c,d}\sum_{n,m}\moy{F_n({\bf x}_c)J_{ij}({\bf x}_a)}\left(\Sigma^{-1}\right)^{cd}_{nm}\moy{F_m({\bf x}_d)J_{kl}({\bf x}_b)}\;.\label{cond_var}
\end{align}
These terms can be evaluated explicitly with the RS ansatz. In particular, using \eqref{inversesig_m}, one finds that the conditional expectation reads:
\begin{align}
\moy{J_{ij}({\bf x}_a)|{\bf F}({\bf x}_b)={\bf m}_b}=&\frac{\Delta_1}{D(2\Delta_0)-D(0)+n\Delta_0 R(2\Delta_0)}\left[(n-1)\delta_{ij}R\left(2\Delta_0\right)+\frac{2}{N}\sum_{b}d_i^{ab}d_j^{ab} R'\left(2\Delta_0\right)\right]\\
&+\frac{\sum_{b}(2D'(2\Delta_0) m_{bi}d_j^{ab}+R(2\Delta_0)m_{bj}d_i^{ab})}{N[D(0)+(n-1)D(2\Delta_0)]}-\frac{D(2 \Delta_0)\sum_{c,b}(2D'(2\Delta_0) \mu_i d_i^{cb}d_j^{ab}+R(2\Delta_0)\mu_j d_j^{cb}d_i^{ab})}{N(D(0)+(n-1)D(2\Delta_0))(D(0)-D(2\Delta_0))} \nn\\
&-\frac{\Delta_1 R(2\Delta_0)\sum_{c,b}(2D'(2\Delta_0) d_i^{cb}d_j^{ab}+R(2\Delta_0)d_j^{cb}d_i^{ab})}{N(D(0)-D(2\Delta_0))(D(0)-D(2\Delta_0)-n \Delta_0 R(2\Delta_0))}\;.\nn
\end{align}
Note that only the first term is of full rank. Following similar steps (see also Ref.~\cite{ros2019complex}), one can show that the conditioning modifies the covariances through \eqref{cond_var} in two ways: by modifying the coefficient of the $\delta_{ij} \delta_{kl}$ term, and by introducing additional terms of finite rank. Therefore, after conditioning the set of $n$ matrices $J({\bf x}_a)$ has joint Gaussian statistics, and the structure of the conditional average and covariance implies that these matrices can be represented via the following decomposition:
\be
J({\bf x}_a)\underset{\rm law}{\equiv} X^a+\xi_a\mathbb{I}+W^a \;.\label{prob_jac}
\ee
Here the $X^a$'s are correlated matrices drawn from the real Elliptic Gaussian Ensemble (rEGE) with
\be\label{eq:StatX}
\moy{X_{ij}^a}=0\;,\;\;\moy{X_{ij}^a X^{b}_{kl}}=\frac{\delta_{jk}\delta_{il}}{N}R\left(d_{ab}^{2}\right)+\frac{2\delta_{ik}\delta_{jl}}{N}D'\left(d_{ab}^{2}\right)\;.
\ee
The random variables $\xi_a$'s are Gaussian, correlated with each other, but independent of the matrices $X^a$'s. We determine their statistics below. Finally, $W^a$ are matrices of  rank $O(n)$ which is correlated to both the matrices $X^a$'s and the random variables $\xi_a$'s. We now argue that the finite rank perturbations do not matter when evaluating the conditional expectation \eqref{eq:CondHess} at leading exponential order in $N$ (within the scaling $k=N\kappa$ with $\kappa>0$). 
As the logarithm of the determinant is a linear statistic, i.e. it only depends on the empirical distribution $\rho_N(z)$ of eigenvalues and not on higher-order correlation functions, and reads
\be
\ln \left|\det {\cal J}({\bf x}_a)\right|={\rm Tr}\ln \left|{\cal J}({\bf x}_a)\right|=N\int d^2 z\,\rho_{a,N}(z)\ln|z+\xi_a|\;,\label{logdet}
\ee
where we denote with $\rho_{a,N}(z)$ the distribution of eigenvalues of the shifted matrices ${\cal J}({\bf x}_a)-\xi_a \mathbb{I}$. This term grows as $O(N)$. One can re-express the conditional expectation above as  
\begin{align}
{\cal E}_{N,n,k}&=\frac{\displaystyle \int \prod_{a=1}^n D\rho_{a,N}\prod_{a=1}^n d\xi_a\,e^{{\cal F}_{N,n}(\{\rho_{a,N}\})+{\cal L}_{N,n}(\{\xi_a\})+N\sum_{a=1}^n \int d^2 z\,\rho_{a,N}(z)\ln|z+\xi_a|}\prod_{a=1}^n \delta_{k,k_a}}{\displaystyle\int \prod_{a=1}^n D\rho_{a,N}\prod_{a=1}^n d\xi_a\,e^{{\cal F}_{N,n}(\{\rho_{a,N}\})+{\cal L}_{N,n}(\{\xi_a\})}}\;,\\
k_a&=N\int_{\Re(z)+\xi_a>0} d^2 z\,\rho_{a,N}(z)\;.\label{eq:IndexShift}
\end{align}
In this equation, ${\cal F}_{N,n}(\{\rho_{a,N}\})$ is the logarithm of the joint distribution of the empirical densities $\rho_{a,N}$, and ${\cal L}_{N,n}(\{\xi_a\})$ corresponds to the logarithm of the Gaussian distribution of the $\{\xi_a\}$. The first term is inherited from the matrices $X^a$'s, that are drawn from the real Elliptic ensemble: it therefore grows as $O(N^2)$. The term ${\cal L}_{N,n}(\{\xi_a\})$ is of order $O(N)$. The finite rank terms are expected to lead to the emergence of a finite number of outliers \cite{EJP3057} but do not impact the bulk limiting distribution of eigenvalues $\rho_a(z)=\lim_{N\to \infty}\rho_{a,N}(z)$. The contribution of these outliers to the linear statistics in \eqref{logdet} is $O(1)$ and thus sub-leading with respect to the $O(N)$ contribution from the bulk. Note however that the contribution from the outliers has to be taken into account when computing the complexity of fixed points of given index $k_a=O(1)$. Here we focus only on $\kappa$-unstable equilibria with $k=N\kappa$, $0<\kappa<1$. For this scaling, one can discard safely the contribution from outliers in the expression of the rescaled index $\kappa_a=\lim_{N\to \infty} k_a/N$, since the extensive index is determined only by the center of the bulk of the eigenvalue distribution, according to \eqref{eq:IndexShift}. The limiting eigenvalue density $\rho_a(z)$ is determined using a saddle-point approximation taking into account only the $O(N^2)$ terms of  ${\cal F}_{N,n}(\{\rho_{a,N}\})$. It corresponds to the optimal density
\be
\rho_a(z)=\rho(z)=\lim_{N\to \infty}\frac{\displaystyle \int \prod_{b=1}^n D\rho_{b,N}\,e^{{\cal F}_{N,n}(\{\rho_{b,N}\})}\rho_{a,N}(z)}{\displaystyle\int \prod_{b=1}^n D\rho_{b,N} \,e^{{\cal F}_{N,n}(\{\rho_{b,N}\})}}=\lim_{N\to \infty}\frac{\displaystyle \int D\rho_{a,N}\,e^{{\cal F}_{N,1}(\rho_{a,N})}\rho_{a,N}(z)}{\displaystyle\int D\rho_{a,N} \,e^{{\cal F}_{N,1}(\rho_{a,N})}}\;,
\ee
which does not depend on the replica index $a$ and corresponds to the marginal density of a matrix $X^a-D^{\mu}$, with $D_{ij}^{\mu}=\delta_{ij}\mu_i$. 

Setting $a=b$ in \eqref{eq:StatX}, one sees that each matrix $X^a$ is drawn from the real Elliptic Gaussian ensemble with parameters
\be
\moy{X_{ij}^a}=0\;,\;\;\moy{X_{ij}^a X^{a}_{kl}}=\frac{c}{N}\left(\delta_{ik}\delta_{jl}+\tau\delta_{il}\delta_{jk}\right)\;,\;\;c=4(\Gamma_V''(0)+\Gamma_A''(0))\;,\;\;\tau=\frac{\Gamma_V''(0)}{\Gamma_V''(0)+\Gamma_A''(0)}\;.
\ee
The limiting distribution $\rho_{c,\tau}(z)$ of the matrix $X^a-D^{\mu}$ has been studied in \cite{lacroix2022counting} (see \cite{sommers1988spectrum} for the case $\mu_i=\mu$) and shown to satisfy 
\be
\int d^2 z\,\rho_{c,\tau}(z)\,\ln |z-x|=-\frac{r(x)}{2c}+\frac{(s(x)-x)^2}{2c\tau}+\frac{1}{2}\int d\mu\,\nu(\mu)\,\ln\left[r(x)+(s(x)+\mu)^2\right]\;,
\ee
which is valid for $x\in [\lambda_-,\lambda_+]$ with $\lambda_{\pm}$ the minimum and maximum real parts of the spectrum, where the functions $r(x)$ and $s(x)$ satisfy 
\be
r(x)=c\int d\mu\,\frac{\nu(\mu)\,r(x)}{r(x)+(s(x)+\mu)^2}\;,\;\;\frac{s(x)-x}{\tau}=-c\int d\mu\,\frac{(\mu+s(x))\,\nu(\mu)}{r(x)+(s(x)+\mu)^2}\;,
\ee
while the index reads
\be
\kappa_a=\int_{\Re(z)+\xi_a>0} d^2 z\,\rho_{c,\tau}(z)=\frac{1}{\pi c\tau}\int_{-\xi_a}^{\infty} dx\,\sqrt{r(x)}\left(1-(1-\tau)s'(x)\right)=\kappa(-\xi_a)\;.
\ee
Let us denote $\lambda(\kappa)$ the inverse function of $\kappa(\lambda)$. In the limit $N\to \infty$ and for an index $0<\kappa<1$, the Kronecker delta constraint $\delta_{N\kappa,N\kappa_a}$ converges up to a sub-leading factor to the delta function $\delta(\xi_a+\lambda(\kappa))$.

The first conclusion we can draw from this analysis is that the determinant term reads, for $k=N\kappa$,
\begin{align}
{\cal E}_{N,n,N\kappa}&\asymp\frac{\displaystyle \int \prod_{a=1}^n d\xi_a\,e^{{\cal L}_{N,n}(\{\xi_a\})+N\sum_{a=1}^n \int d^2 z\,\rho_{c,\tau}(z)\ln|z+\xi_a|}\prod_{a=1}^n \delta\left(\xi_a+\lambda(\kappa)\right)}{\displaystyle\int \prod_{a=1}^n d\xi_a\,e^{{\cal L}_{N,n}(\{\xi_a\})}}\;,
\end{align}
where we define $\asymp$ as
\be
A_N\asymp B_N\;\;\Longleftrightarrow\;\;\lim_{N\to \infty}\frac{1}{N}\ln A_N=\lim_{N\to \infty}\frac{1}{N}\ln B_N\;.
\ee
Using the Gaussian statistics of the random variables $\xi_a$'s, we can express the term ${\cal L}_{N,n}(\{\xi_a\})$ explicitly by introducing the average $\moy{\xi_a}$ and covariance ${\rm Cov}\left(\xi_a,\xi_b\right)=\left(\Sigma_\xi\right)_{ab}/N$. The determinant term can thus be expressed for index $k=N\kappa$ with $0<\kappa<1$ as 
\be
\lim_{N\to \infty}\frac{1}{N}\ln{\cal E}_{N,n,N\kappa}=-\frac{1}{2}\sum_{a,b=1}^n (\lambda(\kappa)+\moy{\xi_a})\left(\Sigma_\xi^{-1}\right)_{ab}(\lambda(\kappa)+\moy{\xi_b})+n\int d^2 z\,\rho_{c,\tau}(z)\,\ln |z-\lambda(\kappa)|\;.
\ee
Importantly the expression above is obtained independently of the hypothesis of replica-symmetry and is thus valid for any number of replica symmetry breaking. It now remains to determine the statistics of the $\xi_a$'s in \eqref{prob_jac}. To perform this calculation explicitly, we rely on the RS assumption. The conditional average reads
\begin{equation}
\langle \xi_a | \mathbf F(\mathbf x_c)=\mathbf m_c\rangle
=\sum_{i=1}^N\sum_{c=1}^n \langle F_i(\mathbf x_c)\xi_a\rangle \sum_{d,m}\left(\Sigma^{-1}\right)^{cd}_{im}m_{dm}
=-\sum_{i=1}^N\sum_{c=1}^n\frac{d^{ac}_i R(d_{ac}^2)}{N}\sum_{d,m}\left(\Sigma^{-1}\right)^{cd}_{im}m_{dm},
\end{equation}
where $\Sigma$ is the covariance matrix \eqref{sigma_matrix}. 
Within the RS ansatz, this term is obtained explicitly from \eqref{inversesig_m}. The conditional covariance of the $\xi_a$'s are given by:
\begin{align}
    {\rm Cov}\left[\xi_a,\xi_b|{\bf F}({\bf x}_c)={\bf m}_{c}\right]&=\moy{\xi_a \xi_b}-\sum_{c,d}\sum_{i,j}\moy{F_i({\bf x}_c)\xi_a}\left(\Sigma^{-1}\right)^{cd}_{ij}\moy{F_j({\bf x}_d)\xi_b}\\
    &=\frac{R\left(d_{ab}^{2}\right)}{N}-\frac{1}{N^2}\sum_{c,d}\sum_{i,j} d_i^{ac}R\left(d_{ac}^{2}\right)\left(\Sigma^{-1}\right)^{cd}_{ij} d_j^{bd} R\left(d_{bd}^{2}\right)\;.
\end{align}
It is convenient to consider the action of the matrix $\Sigma$ on the $3n$-dimensional space formed by the set of $n N$-dimensional vectors $\{\zeta^a,\kappa^a,\chi^a\}$ for $a=1,\cdots,n$ with entries
\be
\zeta_{bi}^a=x_{ai}\;,\;\;\kappa_{bi}^a=\delta_{ab}x_{bi}\;,\;\;\chi_{bi}^a=\delta_{ab}\sum_{c\neq b}x_{ci}\;.
\ee
This set of vectors is closed under the action of $\Sigma$ (or its inverse $\Sigma^{-1}$). One can compute explicitly within the RS ansatz
\begin{align}
    \sum_{b}\sum_j \Sigma_{ij}^{ab}\zeta_{bj}^c&=-\left[D(0)+(n-1)D(2\Delta_0)+\Delta_0 R(2\Delta_0)\right]\zeta_{ai}^c-(n-2)\Delta_0 R(2\Delta_0)\kappa_{ai}^c+\Delta_0 R(2\Delta_0)\chi_{ai}^c+\Delta_0 R(2\Delta_0)\sum_{b\neq c}\kappa_{ai}^b\;,\\
    \sum_{b}\sum_j \Sigma_{ij}^{ab}\kappa_{bj}^c&=-\left[D(2\Delta_0)+\Delta_0 R(2\Delta_0)\right]\zeta_{ai}^c-\left[D(0)-D(2\Delta_0)-\Delta_0 R(2\Delta_0)\right]\kappa_{ai}^c+\Delta_0 R(2\Delta_0)\sum_{b\neq c}\kappa_{ai}^b\;,\\
    \sum_{b}\sum_j \Sigma_{ij}^{ab}\chi_{bj}^c&=\Delta_0 R(2\Delta_0)\zeta_{ai}^c-\Delta_0 R(2\Delta_0)\kappa_{ai}^c-\left[D(0)-D(2\Delta_0)\right]\chi_{ai}^c-D(2\Delta_0)\sum_{b\neq c}\zeta_{ai}^b-\Delta_0 R(2\Delta_0)\sum_{b\neq c}\kappa_{ai}^b\;.
\end{align}
The action of the matrix can thus be expressed within the RS ansatz as
\be
\Sigma \begin{pmatrix}
 \zeta^c\\
 \kappa^c\\
 \chi^c
\end{pmatrix}=A_d\begin{pmatrix}
 \zeta^c\\
 \kappa^c\\
 \chi^c
\end{pmatrix}+A\sum_{b\neq c}\begin{pmatrix}
 \zeta^b\\
 \kappa^b\\
 \chi^b
\end{pmatrix}\;,
\ee
where the matrices $A_d$ and $A$ can be expressed explicitly as
\begin{align}
A_d&=\begin{pmatrix}
 -\left[D(0)+(n-1)D(2\Delta_0)+\Delta_0 R(2\Delta_0)\right] &-(n-2)\Delta_0 R(2\Delta_0)&\Delta_0 R(2\Delta_0) \\
 -\left[D(2\Delta_0)+\Delta_0 R(2\Delta_0)\right]&-\left[D(0)-D(2\Delta_0)-\Delta_0 R(2\Delta_0)\right]&0 \\
 \Delta_0 R(2\Delta_0)&-\Delta_0 R(2\Delta_0)&-\left[D(0)-D(2\Delta_0)\right]
\end{pmatrix}\;,\\
A&=\begin{pmatrix}
 0&\Delta_0 R(2\Delta_0)&0\\
 0&\Delta_0 R(2\Delta_0)&0 \\
 -D(2\Delta_0)& -\Delta_0 R(2\Delta_0)& 0
\end{pmatrix}\;.    
\end{align}
Given the replica-symmetric form of the matrix $\Sigma$, it can be inverted explicitly and satisfies
\be
\Sigma^{-1} \begin{pmatrix}
 \zeta^c\\
 \kappa^c\\
 \chi^c
\end{pmatrix}=(B_d-B)\begin{pmatrix}
 \zeta^c\\
 \kappa^c\\
 \chi^c
\end{pmatrix}+B\sum_{b}\begin{pmatrix}
 \zeta^b\\
 \kappa^b\\
 \chi^b
\end{pmatrix}\;,
\ee
where $B_d$ and $B$ satisfy
\be
B=-\left(A_d-(n-1)A A_d^{-1} A+(n-2)A\right)^{-1}A A_d^{-1}\;,B_d=A_d^{-1}-(n-1)A_d^{-1}A B\;.
\ee  
Thus, within the RS ansatz, one can express
{\small
\begin{align}
&\sum_{c,d}\sum_{n,m} d_n^{ac}R\left(d_{ac}^{2}\right)\left(\Sigma^{-1}\right)^{cd}_{nm} d_m^{bd} R\left(d_{bd}^{2}\right)=R(2\Delta_0)^2\left(\zeta^a-\sum_{c}\kappa^c\right)\Sigma^{-1}\left(\zeta^b-\sum_{d}\kappa^d\right) \\
&=R(2\Delta_0)^2\left(\zeta^a-\sum_{c}\kappa^c\right)\cdot\left[(B_d-B)_{11}\zeta^b+(B_d-B)_{12}\kappa^b+(B_d-B)_{13}\chi^b\right]\nn\\
&-R(2\Delta_0)^2\left(\zeta^a-\sum_{c}\kappa^c\right)\cdot\sum_d\left[\left((B_d+(n-1)B)_{21}-B_{11}\right)\zeta^d+\left((B_d+(n-1)B)_{22}-B_{12}\right)\kappa^d+\left((B_d+(n-1)B)_{23}-B_{13}\right)\chi^d\right]
\nn\;.
\end{align}
}
Using the RS expressions for the dot products
\begin{align}
  &\frac{1}{N}\left(\zeta^a-\sum_{c}\kappa^c\right)\cdot \zeta_b =\Delta_0(n\delta_{ab}-1)\;,\;\;\frac{1}{N}\left(\zeta^a-\sum_{c}\kappa^c\right)\cdot \kappa_b =-\Delta_0(1-\delta_{ab})\;,\\
  &\frac{1}{N}\left(\zeta^a-\sum_{c}\kappa^c\right)\cdot \chi_b =\Delta_0(1-\delta_{ab})\;,
\end{align}
as well as the identity $R\left(d_{ab}^{2}\right)=(R(0)-R(2\Delta_0))\delta_{ab}+R(2\Delta_0)$, one obtains the average and covariance of the Gaussian random variables 
\begin{align}
\moy{\xi_a}&=\bar \xi=\frac{R(2\Delta_0)(n-1)\Delta_1}{D(2\Delta_0)-D(0)+n\Delta_0 R(2\Delta_0)}\;,\;\;{\rm Cov}\left(\xi_a,\xi_b\right)=\frac{\left(\Sigma_\xi\right)_{ab}}{N}\label{av1}\\
\left(\Sigma_\xi\right)_{ab}&=\left[(R(0)-R(2\Delta_0))+n C_n(\Delta_0)\right]\delta_{ab}+R( 2\Delta_0)-\frac{(n-1)\Delta_0 R(2\Delta_0)^2}{D(2\Delta_0)-D(0)+n \Delta_0 R(2\Delta_0)}-C_n(\Delta_0)\;,\label{av2}\\
   C_n(\Delta_0)&=\frac{\Delta_0 R(2\Delta_0)^2 (D(2\Delta_0)-D(0)+\Delta_0 R(2\Delta_0))}{n \Delta_0^2 R(2\Delta_0)^2+\Delta_0 R(2\Delta_0) ((n+1)
   D(2\Delta_0)-D(0))-(D(0)-D(2\Delta_0)) ((n-1)
   D(2\Delta_0)+D(0))}\;.\nn
\end{align}
Note that when $\Gamma_V=0$ (no potential terms in the random force $\tau=0$), $R$ is of $O(1/N)$, see \eqref{CrucialTerms}. The above expressions then imply that the terms $\xi^a$ can be neglected in the calculation of the complexity for $\tau=0$.
Using these expressions, the quadratic form can be expressed within the replica-symmetric ansatz as
\be\label{eq:IndexFix}
\sum_{a,b}\left(\Sigma_\xi^{-1}\right)_{ab}(\lambda(\kappa)+\bar \xi)^2=\frac{n (\lambda(\kappa)+\bar \xi)^2}{\displaystyle R(0)+(n-1)R(2\Delta_0)-\frac{n(n-1)\Delta_0 R(2\Delta_0)^2}{D(2\Delta_0)-D(0)+n \Delta_0 R(2\Delta_0)}}\;.
\ee

Thus, the determinant term can be expressed within the RS ansatz as 
\begin{align}
    \lim_{N\to \infty}\frac{1}{N}\ln{\cal E}_{N,n,N\kappa}=&n\left[-\frac{\displaystyle\left(\lambda(\kappa)+\frac{R(2\Delta_0)(n-1)\Delta_1}{D(2\Delta_0)-D(0)+n\Delta_0 R(2\Delta_0)}\right)^2}{\displaystyle 2\left( R(0)+(n-1)R(2\Delta_0)-\frac{n(n-1)\Delta_0 R(2\Delta_0)^2}{D(2\Delta_0)-D(0)+n \Delta_0 R(2\Delta_0)}\right)}\label{Jac_term}+\int d^2 z\,\rho_{c,\tau}(z)\,\ln |z-\lambda(\kappa)|\right]\;.
\end{align}

\subsection{Expression of the moment at exponential scale}
Gathering all the terms together, one obtains the expression of the factorial moment at exponential scale as 
\begin{align} 
    &\lim_{N\to \infty}\frac{1}{N}\ln{\cal M}^{n}_{k}=\max_\mathbb{Q}\left[\frac{n}{2}\left(1-\int d\mu\, \nu(\mu)\,\ln\left[\sum_{\eta=0}^2\mu^{\eta}\lambda_{\eta}\right]\right)-\frac{1}{2}\int d\mu\,\nu(\mu)\,\ln\left[1+n\left(\frac{\sum_{\eta=0}^2\mu^{\eta}l_{\eta}}{\sum_{\eta=0}^2\mu^{\eta}\lambda_{\eta}}-1\right)\right]\right.\nn\\
    &+n\left[-\frac{\displaystyle\left(\lambda(\kappa)+\frac{R(2\Delta_0)(n-1)\Delta_1}{D(2\Delta_0)-D(0)+n\Delta_0 R(2\Delta_0)}\right)^2}{\displaystyle 2\left( R(0)+(n-1)R(2\Delta_0)-\frac{n(n-1)\Delta_0 R(2\Delta_0)^2}{D(2\Delta_0)-D(0)+n \Delta_0 R(2\Delta_0)}\right)}\right.\nn\\
    &\left.+\int d^2 z\,\rho_{c,\tau}(z)\,\ln |z-\lambda(\kappa)|\right]-\frac{1}{2}\left[n\ln\left(D(2\Delta_0)-D(0)\right)+\ln\left(1-\frac{nD(2\Delta_0)}{D(2\Delta_0)-D(0)}\right)\right]\nn\\
    &+\frac{n}{2}\left(\frac{q_2}{D(0)+(n-1)D(2\Delta_0)}+\frac{(n-1)\Delta_2 D(2\Delta_0)}{(D(0)+(n-1)D(2\Delta_0))(D(0)-D(2\Delta_0))}\right)\nn\\
    &\left.+\frac{n(n-1)\Delta_1^2 R(2\Delta_0) }{2(D(0)-D(2\Delta_0)-n\Delta_0 R(2\Delta_0))(D(0)-D(2\Delta_0))}\right]\nn\;, \\
    \label{RS_moment}
\end{align}
where the maximum over $\mathbb{Q}$ corresponds to a maximum with respect to $Q^0,Q^1,Q^2$, which amounts to a maximum w.r.t $q_0,q_1,q_2,\Delta_0,\Delta_1,\Delta_2$, where 
$\lambda_{\eta}$ and $l_{\eta}$ are given in \eqref{q_lambda}
as functions of these parameters.

Taking the limit $n\to 0$, one obtains
{\small
\begin{align}
    &\lim_{n\to 0}\lim_{N\to \infty}\frac{1}{nN}\ln{\cal M}^{n}_{k}\approx\lim_{N\to \infty}\frac{\moy{\psi({\cal N}_{N,N\kappa}+1)}}{N}\\
    &\approx \max_\mathbb{Q}\left[\frac{1}{2}\left(2-\int d\mu\, \nu(\mu)\,\ln\left[\sum_{\eta=0}^2\mu^{\eta}\lambda_{\eta}\right]\right)-\frac{1}{2}\int d\mu\,\nu(\mu)\,\frac{\sum_{\eta=0}^2\mu^{\eta}l_{\eta}}{\sum_{\eta=0}^2\mu^{\eta}\lambda_{\eta}}-\frac{1}{\displaystyle 2\left( R(0)-R(2\Delta_0)\right)}\left(\lambda(\kappa)-\frac{\Delta_1 R(2\Delta_0)}{D(2\Delta_0)-D(0)}\right)^2\right.\nn\\
    &\left.+\int d^2 z\,\rho_{c,\tau}(z)\,\ln |z-\lambda(\kappa)|-\frac{1}{2}\left[\ln\left(D(2\Delta_0)-D(0)\right)-\frac{D(2\Delta_0)}{D(2\Delta_0)-D(0)}\right]-\frac{1}{2}\left(\frac{q_2}{D(2\Delta_0)-D(0)}+\frac{\Delta_2 D(2\Delta_0)+\Delta_1^2 R(2\Delta_0)}{(D(0)-D(2\Delta_0))^2}\right)\right]\nn\;,
\end{align}
}
with the additional identities
\be
\Delta_{\eta}=\int d\mu\frac{\mu^{\eta}\nu(\mu)}{\sum_{\omega=0}^2 \mu^{\omega}\lambda_{\omega}}\;,\;\;q_{\eta}=2\Delta_{\eta}-\int d\mu\frac{\mu^{\eta}\nu(\mu)}{\left(\sum_{\omega=0}^2 \mu^{\omega}\lambda_{\omega}\right)^2}\sum_{\omega=0}^2 \mu^{\omega}l_{\omega}\;,\;\;\eta=0,1,2\;.
\ee
Note that $\Delta_{\eta}$ is independent of the $l_{\omega}$'s while $q_{\eta}$ depends on them explicitly. The optimisation can thus be taken with respect to the $\Delta_{\eta}$'s and $l_{\eta}$'s by replacing $q_2$ by its expression above. Considering the optimisation with respect to the $l_\eta$'s turns out to be sufficient to obtain the expression of the quantity above. The saddle-point equations read
\be
\int d\mu\frac{\mu^{\eta}\nu(\mu)}{\left(\sum_{\omega=0}^2 \mu^{\omega}\lambda_{\omega}\right)^2}\left[\sum_{\omega=0}^2 \mu^{\omega}\lambda_{\omega}-\frac{\mu^2}{D(2\Delta_0)-D(0)}\right]=0\;,
\ee
which thus all share the simple solution $\sum_{\omega=0}^2 \mu^{\omega}\lambda_{\omega}=\mu^2/[D(2\Delta_0)-D(0)]$, yielding the simple expressions
\be
\Delta_{\eta}=(D(2\Delta_0)-D(0))\int d\mu\frac{\nu(\mu)}{\mu^{2-\eta}}=(D(2\Delta_0)-D(0))I_{2-\eta}\;.
\ee
Using these identities, one obtains 
\begin{align}
    \lim_{N\to \infty}\frac{\moy{\psi({\cal N}_{N,N\kappa}+1)}}{N}
   =&\frac{1}{2}\left(2-\int d\mu\, \nu(\mu)\,\ln\frac{\mu^2}{D(2\Delta_0)-D(0)}\right)-\frac{\displaystyle\left(\lambda(\kappa)- I_1 R(2\Delta_0)\right)^2}{\displaystyle 2\left( R(0)-R(2\Delta_0)\right)}+\frac{1}{2}\left(-2-\frac{D(2\Delta_0)}{D(2\Delta_0)-D(0)}\right)\nn\\
    &+\int d^2 z\,\rho_{c,\tau}(z)\,\ln |z-\lambda(\kappa)|-\frac{1}{2}\left[\ln\left(D(2\Delta_0)-D(0)\right)-\frac{D(2\Delta_0)}{D(2\Delta_0)-D(0)}\right]-\frac{\displaystyle  I_1^2 R(2\Delta_0)}{\displaystyle 2}\nn\\
    =&\int d^2 z\,\rho_{c,\tau}(z)\,\ln |z-\lambda(\kappa)|-\int d\mu\, \nu(\mu)\,\ln \mu-\frac{\displaystyle\left(\lambda(\kappa)-I_1 R(2\Delta_0)\right)^2}{\displaystyle 2\left( R(0)-R(2\Delta_0)\right)}-\frac{\displaystyle I_1^2 R(2\Delta_0)}{\displaystyle 2}\;.
\end{align}
We now specify to the case of homogeneous rates $\mu_i=\mu$. In this case $\lambda(\kappa)=\sqrt{c}(1+\tau)z(\kappa)-\mu$ and for any $0<\kappa<1$, this expression can be simplified as
\be
\int d^2 z\,\rho_{c,\tau}(z)\,\ln |z-\lambda(\kappa)|=\frac{(\lambda(\kappa)+\mu)^2}{2c(1+\tau)}-\frac{1}{2}+\frac{1}{2}\ln c=\frac{1+\tau}{2}z(\kappa)^2-\frac{1}{2}+\frac{1}{2}\ln c\;.
\ee
The value of $\Delta_0$ is determined from $\Delta_0=( D(2\Delta_0)-D(0))/\mu^2$. As described in the main text, the complexity does not depend independently on $c=4\Gamma''(0)$ and $\mu$ but only on the ratio $\gamma=c/\mu^2$. For simplicity, we also consider the case where
\be
\Gamma_V(q)=\tau\Gamma(q)\;,\;\;\Gamma_A(q)=(1-\tau)\Gamma(q)\;.
\ee
We thus introduce the scaling functions 
\be
\Delta_0=\frac{1}{2}{\cal Q}\left(\gamma=\frac{c}{\mu^2}\right)\;,\;\;R(2\Delta_0)=R(0)v\left(\gamma=\frac{c}{\mu^2}\right)=\mu^2\gamma\tau v(\gamma)\;,
\ee
where we have used that $ R(0)=c\tau=\mu^2\gamma\tau$. In the large $N$ limit, $R(q)=4\Gamma_V''(q)$ such that it is a positive, decreasing function of $q$. As a result, $0\leq v(\gamma)\leq 1$. Using now that $D(q)=2\Gamma'(q)$ at leading order in the large $N$ limit, we can re-express the saddle-point condition on $\Delta_0$ as a self-consistent equation for the associated scaling function ${\cal Q}(\gamma)$ which reads 
\be
\gamma=\frac{4\Gamma''(0)}{\mu^2}=\frac{4\Gamma''(0)\Delta_0}{D(2\Delta_0)-D(0)}=\frac{{\cal Q}(\gamma)\Gamma''(0)}{\Gamma'({\cal Q}(\gamma))-\Gamma'(0)}\;.
\ee
One finally obtains
\begin{align}
    \lim_{N\to \infty}\frac{\moy{\psi({\cal N}_{N,N\kappa}+1)}}{N}
    =&\frac{1+\tau}{2}z(\kappa)^2-\frac{\tau\gamma v(\gamma)}{2}-\frac{1}{2}+\frac{1}{2}\ln \gamma-\frac{1}{\displaystyle 2\tau\left( 1-v\left(\gamma\right)\right)}\left((1+\tau)z(\kappa)-\frac{1}{\sqrt{\gamma}}-\tau\sqrt{\gamma} v\left(\gamma\right)\right)^2\;,
    \label{Xi_kappa}
\end{align}
where $-1<z(\kappa)<1$ satisfies
\be
\kappa=\int_{\Re(z)\geq \sqrt{c}(1+\tau)z(\kappa)} \frac{d^2z}{\pi c(1-\tau^2)}\Theta\left(c-\frac{\Re(z)^2}{(1+\tau)^2}-\frac{\Im(z)^2}{(1-\tau)^2}\right)=\frac{1}{\pi}\left[{\rm arccos}\left(z(\kappa)\right)-z(\kappa)\sqrt{1-z(\kappa)^2}\right]\;.
\ee
Finally, to obtain the statistics of ${\cal N}_{N,\rm tot}$, one needs to optimise with respect to $\kappa$ (or alternatively $z(\kappa)$), yielding
\begin{align}
    \lim_{N\to \infty}\frac{\moy{\psi({\cal N}_{N,\rm tot}+1)}}{N}
    =\lim_{N\to \infty}\frac{\moy{\ln {\cal N}_{N,\rm tot}}}{N}=&\frac{1}{2}\left(\ln \gamma-\frac{\displaystyle\left(\gamma-1\right)\left(1+\tau\gamma v\left(\gamma\right)\right)}{\displaystyle \gamma\left( 1+\tau v\left(\gamma\right)\right)}\right)\;,
\end{align}
where at the optimal value $\kappa_*$, one has
\be
z(\kappa_*)=\frac{\displaystyle 1+\tau\gamma v\left(\gamma\right)}{\displaystyle \sqrt{\gamma}(1+\tau v\left(\gamma\right))}\;.
\ee

Recall that throughout the derivation we have worked with the rescaled variables introduced in \eqref{c1}-\eqref{c3} and, for notational convenience, dropped the hats. All definitions and equations above should therefore be understood in terms of these rescaled variables. Conveniently, the final expression for the complexity depends only on the dimensionless quantities \(\gamma,\tau,\kappa,z(\kappa)\), and \(v(\gamma)\), which are invariant under this rescaling. To make contact with the original convention, we temporarily restore the hats for the rescaled quantities. The original variables are related to them through
\be
\lambda=\sqrt{N}\,\hat\lambda,\qquad
\mu=\sqrt{N}\,\hat\mu,\qquad
c=N\hat c
=4N\big[\Gamma_V''(0)+\Gamma_A''(0)\big].
\ee
In particular,
\be
\gamma=\frac{\hat c}{\hat\mu^2}
=\frac{c}{\mu^2},
\qquad
\tau=
\frac{\Gamma_V''(0)}
{\Gamma_V''(0)+\Gamma_A''(0)}.
\ee
Thus the final expressions for the complexity are identical in the original and rescaled conventions.

\section{Stability of the replica symmetric solution}
After obtaining the expression of the quenched complexity within the replica symmetric ansatz, the next natural step is to investigate the stability of that solution. We adapt to our model the procedure of \cite{AT78}. To this end, we need to express the expression of the factorial moments for arbitrary $n$ in a general form without relying on replica-symmetry. The expression of the volume element is straightforward
\begin{align}
    \lim_{N\to \infty}\frac{1}{N}\ln{\cal V}_{N,n}(\mathbb{Q})=\frac{1}{2}\min_{L^0,L^1,L^2}\left[{\rm Tr}\left(\sum_{\eta=0}^2 L^{\eta}Q^{\eta}\right)-\int d\nu(\mu)\,\ln \det\left(\sum_{\eta=0}^2\mu^{\eta}L^{\eta}\right)+n\ln(2\pi)\right]\;.
\end{align}
For the distribution of fields and the Jacobian terms, it will be useful to introduce the following notations and identities
\begin{align}
    -\frac{1}{2}\left[\frac{1}{N}{\bf m}^T\Sigma^{-1}{\bf m}+(\lambda{\bf 1}+\bar \xi)^T\Sigma_{\xi}^{-1}(\lambda{\bf 1}+\bar \xi)\right]=-\frac{1}{2}\begin{pmatrix}
        \frac{{\bf m}^T}{\sqrt{N}}&-\lambda{\bf 1}^T
    \end{pmatrix}{\cal C}^{-1}\begin{pmatrix}
        \frac{{\bf m}}{\sqrt{N}}\\ -\lambda{\bf 1}
    \end{pmatrix}\;,
\end{align}
where the matrix ${\cal C}$ has the following block decomposition
\begin{align}
{\cal C}&=\begin{pmatrix}
    \Sigma_{FF}&\Sigma_{F\xi}\\
    \Sigma_{\xi F}&\Sigma_{\xi\xi}
\end{pmatrix}\;,\\
\Sigma_{FF;ai,bj}&=\moy{F_i({\bf x}_a)F_j({\bf x}_b)}=-\delta_{ij}D(d_{ab}^2)-\frac{d_i^{ab}d_j^{ab}}{N}R(d_{ab}^2)\\
\Sigma_{F\xi;ai,b}&=\sqrt{N}\moy{F_i({\bf x}_a)\xi_b}=\frac{d_i^{ab}}{\sqrt{N}}R(d_{ab}^2)\;,\;\;\Sigma_{\xi\xi;a,b}=N\moy{\xi_a\xi_b}=R(d_{ab}^2)\;.
\end{align}
The term above can be conveniently rewritten as
\begin{align}
&-\frac{1}{2}\begin{pmatrix}
        \frac{{\bf m}^T}{\sqrt{N}}&-\lambda{\bf 1}^T
    \end{pmatrix}{\cal C}^{-1}\begin{pmatrix}
        \frac{{\bf m}}{\sqrt{N}}\\ -\lambda{\bf 1}
    \end{pmatrix}=\min_{{\bf t},{\bf p}}\left[\frac{1}{2}\begin{pmatrix}
        \frac{{\bf p}^T}{\sqrt{N}}&{\bf t}^T
    \end{pmatrix}{\cal C}\begin{pmatrix}
       \frac{{\bf p}}{\sqrt{N}}\\ {\bf t}
    \end{pmatrix}-\begin{pmatrix}
        \frac{{\bf p}^T}{\sqrt{N}}&{\bf t}^T
    \end{pmatrix}\begin{pmatrix}
        \frac{{\bf m}}{\sqrt{N}}\\ -\lambda{\bf 1}
    \end{pmatrix}\right]\\
    =&\min_{{\bf t},{\bf p}}\left[-\frac{1}{2}\sum_{a,b}\left[D(d_{ab}^2)\frac{{\bf p}_a\cdot {\bf p}_b}{N}+R(d_{ab}^2)\left(\frac{{\bf p}_a\cdot {\bf d}^{ab}}{N}\right)\left(\frac{{\bf p}_b\cdot {\bf d}^{ab}}{N}\right)-2R(d_{ab}^2)\left(\frac{{\bf p}_a\cdot {\bf d}^{ab}}{N}\right)t_b-R(d_{ab}^2)t_a t_b\right]\right.\\
    &\left.-\sum_a \left(\frac{{\bf p}_a\cdot {\bf m}_a}{N}-\lambda t_a\right)\right] 
\end{align}
At this stage, we introduce the overlap matrices
\be
P_{ab}=\frac{{\bf p}_a\cdot {\bf p}_b}{N}\;,\;\;C_{ab}^0=\frac{{\bf p}_a\cdot {\bf x}_b}{N}\;,\;\;C_{ab}^1=\frac{{\bf p}_a\cdot {\bf m}_b}{N}\;,\tilde D_{ab}=-D(d_{ab}^2)\;,\;R_{ab}=R(d_{ab}^2)\;.
\ee
The matrices $P,C^0,C^1$ are not independent as one needs to optimise over the values of the ${\bf p}_a$'s. Let us decompose ${\bf p}_{a}$ into a component ${\bf p}_{a}^{\perp}$ orthogonal to all the $2n$ vectors ${\cal S}_{2n}=\{{\bf x}_a,{\bf m}_a\}$ and a component ${\bf p}_{a}^{\parallel}$ in the span ${\cal S}_{2n}$. Note that for a fixed $\mu_i=\mu$, ${\bf m}_a=\mu{\bf x}_a$ are colinear and one can restrict to the span of the ${\bf x}_a$'s only. For the orthogonal component, the minimisation is over $\frac{1}{2N}\sum_{a,b}\tilde D_{ab}{\bf p}_a^{\perp}{\bf p}_b^{\perp}$. As the matrix $\tilde D_{ab}$ is positive definite, the minimum is obtained for ${\bf p}_a^{\perp}={\bf 0}$. Thus ${\bf p}_{a}$ can be decomposed onto a basis of ${\cal S}_{2n}$. The basis can be written as 
\be
\begin{pmatrix}
    {\bf e}^0\\{\bf e}^1
\end{pmatrix}=\frac{1}{\sqrt{N}}\begin{pmatrix}
    Q^0&Q^1\\
    Q^1&Q^2
\end{pmatrix}^{-1/2}\begin{pmatrix}
    {\bf x}\\{\bf m}
\end{pmatrix}
\ee
Expanding the ${\bf p}_a$'s in this basis, one can write $P$ explicitly as
\be
P=\begin{pmatrix}
    C^0&C^1
\end{pmatrix}\begin{pmatrix}
    Q^0&Q^1\\
    Q^1&Q^2
\end{pmatrix}^{-1}\begin{pmatrix}
    {C^0}^T\\{C^1}^T
\end{pmatrix}\;.
\ee
It yields
\begin{align}
-\frac{1}{2}\begin{pmatrix}
        \frac{{\bf m}^T}{\sqrt{N}}&-\lambda{\bf 1}^T
    \end{pmatrix}{\cal C}^{-1}\begin{pmatrix}
        \frac{{\bf m}}{\sqrt{N}}\\ -\lambda{\bf 1}
    \end{pmatrix}
    =\min_{{\bf t},C^0,C^1}\left[\frac{1}{2}\sum_{a,b}\left[\tilde D_{ab}P_{ab}+R_{ab}\left(C_{aa}^0-C_{ab}^0+t_a\right)\left(C_{bb}^0-C_{ba}^0+t_b\right)\right]-\sum_a \left(C_{aa}^1-\lambda t_a\right)\right]\;, 
\end{align}
with $P$ as indicated above. Gathering all the terms of the action, one finally obtains
\begin{align}
    \Phi_n[\mathbb{Q}]=&\frac{1}{2}\min_{L^0,L^1,L^2}\left[{\rm Tr}\left(\sum_{\eta=0}^2 L^{\eta}Q^{\eta}\right)-\int d\nu(\mu)\,\ln \det\left(\sum_{\eta=0}^2\mu^{\eta}L^{\eta}\right)\right]-\frac{1}{2}\ln \det \tilde D+n\int d^2 z\,\rho_{c,\tau}(z)\,\ln |z-\lambda|\\
    &+\min_{{\bf t},C^0,C^1}\left[\frac{1}{2}\sum_{a,b}\left[\tilde D_{ab}P_{ab}+R_{ab}\left(C_{aa}^0-C_{ab}^0+t_a\right)\left(C_{bb}^0-C_{ba}^0+t_b\right)\right]-\sum_a \left(C_{aa}^1-\lambda t_a\right)\right]\;.
\end{align}
The stability of this action can be analysed for arbitrary limiting distribution $\nu(\mu)$ but for simplicity, we focus in the following on the case of fixed $\mu$. The action simplifies in that case and reads
\begin{align}
    \Phi_n[\mathbb{Q}]=&\frac{1}{2}\left[n+\ln \det Q-\ln \det \tilde D\right]+n\int d^2 z\,\rho_{c,\tau}(z)\,\ln |z-\lambda|\\
    &+\min_{{\bf t},C}\left[\frac{1}{2}\sum_{a,b}\left[\tilde D_{ab}\left(C Q^{-1}C^T\right)_{ab}+R_{ab}\left(C_{aa}-C_{ab}+t_a\right)\left(C_{bb}-C_{ba}+t_b\right)\right]-\sum_a \left(\mu C_{aa}-\lambda t_a\right)\right]\;.
\end{align}
With this expression, we are now able to perform the following expansion
\begin{align}
    Q=Q_{\rm RS}+\epsilon H\;,\;\;C=C_{\rm RS}+\epsilon K\;,\;\;{\bf t}=t_0{\bf 1}+\epsilon {\bf u}\;,
\end{align}
up to second order in $\epsilon$. It yields 
\begin{align}
\sigma_{ab}=H_{aa}+H_{bb}-2H_{ab}\;,\;\;\chi_{ab}=K_{aa}-K_{ab}+u_a\;,\;\;X_{ab}=C_{{\rm RS},aa}-C_{{\rm RS},ab}+t_0\;.
\end{align}
The expansion reads
\be
\Phi_n=\Phi_n^{(0)}+\epsilon \Phi_n^{(1)}+\frac{\epsilon^2}{2} \Phi_n^{(2)}+o(\epsilon^2)\;,
\ee
where the zeroth order term reads
\begin{align}
 \Phi_n^{(0)}=&\frac{n}{2}\left[1-\ln\frac{D(2\Delta)-D(0)}{\Delta}\right]+\frac{1}{2}\left[\ln\left(1+n\left(\frac{q}{\Delta}-1\right)\right)-\ln\left(1-n\frac{D(2\Delta)}{D(2\Delta)-D(0)}\right)\right]+n\int d^2 z\,\rho_{c,\tau}(z)\,\ln |z-\lambda|\\
    &+\frac{n-1}{2}\left(D(2\Delta)-D(0)\right)\frac{\Delta_C^2}{\Delta}+\frac{1}{2}\left((1-n)D(2\Delta)-D(0)\right)\frac{((1-n)\Delta_C+nC_0)^2}{((1-n)\Delta+n q)}\\
    &+\frac{n}{2}R(0)t_0^2+\frac{n(n-1)}{2}R(2\Delta)(t_0+\Delta_C)^2-n\left(\mu C_0-\lambda t_0\right)\;,
\end{align}
where we have used the notation $C_{{\rm RS},ab}=C_0-\Delta_C(1-\delta_{ab})$. The first order term reads instead
\begin{align}
 \Phi_n^{(1)}=&\frac{1}{2}{\rm Tr}\left[Q_{\rm RS}^{-1}H-\left(\tilde D_{\rm RS}\right)^{-1}\delta\tilde D\right]-\sum_a \left(\mu K_{aa}-\lambda u_a\right)\\
 &+\frac{1}{2}\sum_{a,b}\left[\delta\tilde D_{ab}P_{\rm RS,ab}+\tilde D_{{\rm RS},ab}\delta P_{ab}+\delta R_{ab}X_{ab}X_{ba}+ R_{{\rm RS},ab}X_{ab}\chi_{ba}+R_{{\rm RS},ab}X_{ba}\chi_{ab}\right]\;,
\end{align}
with 
\begin{align}
    &\delta \tilde D_{ab}=-D'(2\Delta(1-\delta_{ab}))\sigma_{ab}\;,\;\;\tilde D_{{\rm RS},ab}=-D(2\Delta(1-\delta_{ab}))\;,\\
    &\delta R_{ab}=R'(2\Delta(1-\delta_{ab}))\sigma_{ab}\;,\;\;R_{{\rm RS},ab}=R(2\Delta(1-\delta_{ab}))\;,\\
    &\delta P=K Q_{\rm RS}^{-1}C_{\rm RS}^T+C_{\rm RS} Q_{\rm RS}^{-1}K^T-C_{\rm RS}Q_{\rm RS}^{-1}H Q_{\rm RS}^{-1}C_{\rm RS}^T\;.
\end{align}
Finally, by conveniently grouping terms together, one obtains the following expression
\begin{align}
\Phi_n^{(1)}=&A_{n,Q}^d\sum_{a}H_{aa}+A_{n,Q}^0\sum_{a\neq b}H_{ab}+A_{n,C}^d\sum_{a}K_{aa}+A_{n,C}^0\sum_{a\neq b}K_{ab}+A_{n,t}\sum_a u_a\;,\\
 A_{0,Q}^d=&\frac{1}{2\Delta}\left[1-\Delta_C^2\frac{D(2\Delta)-D(0)}{\Delta}\right]+A_{0,Q}^0\;,\\
 A_{0,Q}^0=&-\frac{q-\Delta}{2\Delta^2}+\frac{D(2\Delta)\Delta_C^2}{2\Delta^2}+\frac{D(2\Delta)-D(0)}{\Delta}\left[\frac{\Delta_C^2(q-\Delta)}{\Delta^2}-\frac{\Delta_C(C_0-\Delta_C)}{\Delta}\right]-R'(2\Delta)(t_0+\Delta_C)^2\\
 &+D'(2\Delta)\left[\frac{2\Delta_C(C_0-\Delta_C)}{\Delta}-\frac{\Delta_C^2(q-\Delta)}{\Delta^2}-\frac{D(2\Delta)}{(D(2\Delta)-D(0))^2}\right]\;,\\
 A_{0,C}^d=&\frac{D(2\Delta)-D(0)}{\Delta}\Delta_C-\mu+A_{0,C}^0\;,\\
 A_{0,C}^0=&\frac{D(2\Delta)-D(0)}{\Delta}\left[C_0-\frac{\Delta_C q}{\Delta}\right]-\frac{\Delta_C}{\Delta}D(2\Delta)-R(2\Delta)(t_0+\Delta_C)\;,\\
 A_{0,t}=&(R(0)-R(2\Delta))t_0-R(2\Delta)\Delta_C+\lambda\;.
\end{align}
Setting all the five coefficients to zero, one recovers the saddle-point equations and in particular
\begin{align}
    \Delta_C=\frac{1}{\mu}\;,\;\;\mu^2=\frac{D(2\Delta)-D(0)}{\Delta}\;,\;\;t_0=\frac{R(2\Delta)-\mu\lambda}{\mu(R(0)-R(2\Delta))}\;.
\end{align}
Taking the limit $\lim_{n\to 0}\Phi_n^{(0)}/n$ and inserting the expressions above, one recovers
\be
\lim_{n\to 0}\frac{\Phi_n^{(0)}}{n}=\lim_{N\to \infty}\frac{\moy{\psi({\cal N}_{N,N\kappa}+1)}}{N}=\int d^2 z\,\rho_{c,\tau}(z)\,\ln |z-\lambda|-\ln \mu-\frac{\left(\displaystyle \frac{R(2\Delta)}{\mu}-\lambda\right)^2}{2[R(0)-R(2\Delta)]}-\frac{R(2\Delta)}{2\mu^2}\;.
\ee
Finally, let us consider the second order term. It reads
\begin{align}
 \Phi_n^{(2)}=&-\frac{1}{2}{\rm Tr}\left[Q_{\rm RS}^{-1}H Q_{\rm RS}^{-1}H-\left(\tilde D_{\rm RS}\right)^{-1}\delta\tilde D\left(\tilde D_{\rm RS}\right)^{-1}\delta\tilde D+\left(\tilde D_{\rm RS}\right)^{-1}\delta^2\tilde D\right]\\
 &+\frac{1}{2}\sum_{a,b}\left[\delta^2\tilde D_{ab}P_{\rm RS,ab}+\tilde D_{{\rm RS},ab}\delta^2 P_{ab}+2\delta \tilde D_{ab}\delta P_{ab}+\delta^2 R_{ab}X_{ab}X_{ba}+ 2\delta R_{ab}(X_{ab}\chi_{ba}+X_{ba}\chi_{ab})+2R_{{\rm RS},ab}\chi_{ba}\chi_{ab}\right]\;,
\end{align}
with 
\begin{align}
    &\delta^2 \tilde D_{ab}=-D''(2\Delta(1-\delta_{ab}))\sigma_{ab}^2\;,\delta^2 R_{ab}=R''(2\Delta(1-\delta_{ab}))\sigma_{ab}^2\;,\\
    &\delta^2 P=2K Q_{\rm RS}^{-1}K^T-2KQ_{\rm RS}^{-1}HQ_{\rm RS}^{-1}C_{\rm RS}^T-2C_{\rm RS} Q_{\rm RS}^{-1}HQ_{\rm RS}^{-1}K^T+2C_{\rm RS}Q_{\rm RS}^{-1}H Q_{\rm RS}^{-1}H Q_{\rm RS}^{-1}C_{\rm RS}^T\;.
\end{align}
The term $\Phi_n^{(2)}$ can be rewritten as a quadratic form
\be
\Phi_n^{(2)}=H^T {\cal A}_1 H+2H^T {\cal A}_2 Y+Y^T {\cal A}_3 Y\;,\;\;Y=(K,{\bf u})\;.
\ee
As the expression of interest is optimised over the $Y$ sector, we focus on the effective $H$ sector after optimisation, which reads
\be
\Phi_{n,{\rm eff}}^{(2)}=H^T{\cal M} H\;,\;\;{\cal M}={\cal A}_1-{\cal A}_2{\cal A}_3^{-1}{\cal A}_2^T\;.
\ee
The RS saddle is invariant under permutations of the replicas. The quadratic form decomposes into longitudinal, anomalous and replicon sectors. Similarly $H,K,{\bf u}$ can be decomposed into these sectors and different sectors do not couple with each other. The optimisation over $Y=(K,{\bf u})$ can be done sector by sector without constructing explicitly the full matrix ${\cal M}$. 

For the replicon sector, decomposing $H=h{\cal R}$, $K=k{\cal R}$ and ${\bf u}={\bf 0}$ onto a replicon matrix ${\cal R}$ such that
\be
{\cal R}_{ab}={\cal R}_{ba}\;,\;\;{\cal R}_{aa}=0\;,\;\;\sum_b {\cal R}_{ab}=0\;,\;\;\sum_{ab}{\cal R}_{ab}^2=1\;,
\ee
one obtains
\begin{align}
\Phi_{0,r}^{(2)}=&a_r h^2+2b_r h k +c_r k^2\;,\\
a_r=&\frac{(2D'(2\Delta)-\mu^2)^2}{2\mu^4\Delta^2}+2\left(\frac{R(0)-\lambda\mu}{\mu(R(0)-R(2\Delta))}\right)^2\left[R''(2\Delta)-\frac{2R'(2\Delta)D''(2\Delta)}{2D'(2\Delta)-\mu^2}\right]\\
b_r=&\frac{2D'(2\Delta)-\mu^2}{\mu\Delta}+2R'(2\Delta)\frac{R(0)-\lambda\mu}{\mu(R(0)-R(2\Delta))}\;,\;\;c_r=\mu^2+R(2\Delta)
\end{align}
Optimising over $k$, one obtains explicitly $k_*=-b_r/c_r h$ and the replicon eigenvalue
\begin{align}
\Lambda_r=a_r-\frac{b_r^2}{c_r}=&\frac{(2D'(2\Delta)-\mu^2)^2}{2\mu^4\Delta^2}+2\left(\frac{R(0)-\lambda\mu}{\mu(R(0)-R(2\Delta))}\right)^2\left[R''(2\Delta)-\frac{2R'(2\Delta)D''(2\Delta)}{2D'(2\Delta)-\mu^2}\right]\\
&-\frac{1}{\mu^2+R(2\Delta)}\left[\frac{2D'(2\Delta)-\mu^2}{\mu\Delta}+2R'(2\Delta)\frac{R(0)-\lambda\mu}{\mu(R(0)-R(2\Delta))}\right]^2\;.\nonumber    
\end{align}
The expression above can be rewritten in terms of the dimensionless variables $\gamma,\tau,v(\gamma),z(\kappa)$ as
\begin{align}
\Lambda_r&=\frac{2(1-\gamma v(\gamma))^2}{{\cal Q}(\gamma)^2}\left[1+\frac{\gamma A^2}{\tau(1-v(\gamma))^2}\left(v''(\gamma)-\frac{2v(\gamma)v'(\gamma)}{1-\gamma v(\gamma)}\right)-\frac{2}{1+\tau\gamma v(\gamma)}\left(1-\frac{A\gamma}{1-v(\gamma)}v'(\gamma)\right)^2\right]\;,\\
A&=1+\tau\gamma-\sqrt{\gamma}(1+\tau)z(\kappa)\;.
\end{align}
When evaluating this expression for the total complexity, one should consider $z(\kappa_*)$, and thus $A=A_*$ with
\be
A_*=\frac{\tau(\gamma-1)(1-v(\gamma))}{1+\tau v(\gamma)}\;.
\ee
In the limit $\tau\to 0$, one can immediately check that
\be
\Lambda_r^{\rm tot}=-\frac{2(1-\gamma v(\gamma))^2}{{\cal Q}(\gamma)^2}\leq 0\;,
\ee
and the replicon predicts stability of the RS ansatz for any $\gamma>1$. In addition, one can expand the expression of the replicon in the vicinity of the transition, i.e. $\gamma\to 1$ for arbitrary $0\leq \tau\leq 1$, yielding 
\be
\Lambda_r^{\rm tot}=-\frac{1}{2}\left(\frac{\Gamma'''(0)}{\Gamma''(0)}\right)^2\left[\frac{1-\tau}{1+\tau}+2\left[\left(1-\frac{4\Gamma''(0)\Gamma''''(0)}{3\Gamma'''(0)^2}\right)\frac{1-\tau}{1+\tau}+\frac{3\tau}{(1+\tau)^2}\right](\gamma-1)+O(\gamma-1)^2\right]\;.
\ee
Thus, one can check that the replicon predicts stability of the RS ansatz directly above the transition for arbitrary $0\leq \tau\leq 1$. Thus, we expect the expansions of the complexities based on the RS ansatz in the parameter $\gamma-1$ to be the correct expansion of the complexities.

For the anomalous sector, consider a vector ${\bf v}$ such that
\be
\sum_a v_a=0\;,\;\;\sum_a v_a^2=1\;.
\ee
We then decompose the quadratic form using
\begin{align}
    &H_{aa}=h_d v_a\;,\;\;H_{ab}=h_0(v_a+v_b)\;,\;\;u_a=u v_a\;,\\
    &K_{aa}=k_d v_a\;,\;\;K_{ab}=k_1 v_a+k_2 v_b\;,\;\;a\neq b\;.
\end{align}
Denoting ${\bf h}^T=\begin{pmatrix}
    h_d&h_0
\end{pmatrix}$ and ${\bf y}^T=\begin{pmatrix}
    k_d&k_1&k_2&u
\end{pmatrix}$, the quadratic form for the anomalous sector reads
\begin{align}
\Phi_{0,a}^{(2)}=&{\bf h}^T A_{a}{\bf h}+2{\bf h}^T B_a{\bf y} +{\bf y}^T C_a {\bf y}\;,
\end{align}
The optimisation yields ${\bf y}_*=-C_a^{-1}B_a^T {\bf h}$. Thus, one should consider the eigenvalues of 
\be
M_a=A_a-B_a C_a^{-1}B_a^T\;.
\ee
At finite $n$, the norm of vectors in the anomalous sector reads
\be
\sum_{ab}H_{ab}^2=h_d^2+2(n-2)h_0^2\;,
\ee
thus, in the limit $n\to 0$, the eigenvalues for this sector are solutions of the generalised problem
\be
\det\left(\begin{pmatrix}
    1&0\\
    0&-4
\end{pmatrix}^{-1}M_a-\Lambda\mathbb{I}\right)=0\;.
\ee
The product of matrices can be expressed as
\be
\begin{pmatrix}
    1&0\\
    0&-4
\end{pmatrix}^{-1}M_a=\Lambda_a \mathbb{I}+\Xi_a\begin{pmatrix}
    1&-2\\
    1/2&-1
\end{pmatrix}\;,
\ee
where the second matrix is nilpotent. Thus, the two eigenvalues of the anomalous sector coincide and read
\be
\Lambda=\Lambda_a=\frac{2D'(2\Delta)-\mu^2}{2\mu^2 \Delta^2}=-\frac{2(1-\gamma v(\gamma))}{{\cal Q}(\gamma)^2}\leq 0\;,
\ee
which using the identity
\be
\frac{1}{\gamma}=\frac{1}{{\cal Q}(\gamma)}\int_0^{{\cal Q}(\gamma)}\frac{\Gamma''(s)}{\Gamma''(0)}ds\geq v(\gamma)=\frac{\Gamma''({\cal Q}(\gamma))}{\Gamma''(0)}\;,
\ee
shows that the anomalous sector is stable.

Finally, for the longitudinal sector, one should take
\begin{align}
    H_{aa}=h_d\;,\;\;H_{ab}=h_0\;,\;\;u_a=u\\
    K_{aa}=k_d\;,\;\;K_{ab}=k_0\;,\;\;a\neq b\;,
\end{align}
which corresponds to the usual RS parameterisation. Optimising over the auxiliary variables $K,{\bf u}$, the equations for the longitudinal sector are similar to the generalised eigenvalue problem for the anomalous sector. At finite $n$, the norm of vectors in the longitudinal sector reads
\be
\sum_{ab}H_{ab}^2=n h_d^2+n(n-1)h_0^2\;,\;\lim_{n\to 0 }\frac{1}{n}\sum_{ab}H_{ab}^2=h_d^2-h_0^2\;,
\ee
and in the limit $n\to 0$, one should solve the generalised problem 
\be
\det\left(
    \begin{pmatrix}
    1&0\\
    0&-1
\end{pmatrix}^{-1}M_l-\Lambda\mathbb{I}\right)=\det\left((\Lambda_l-\Lambda)\mathbb{I}+\Xi_l\begin{pmatrix}
    1&-1\\
    1&-1
\end{pmatrix}\right)=0\;.
\ee
One can actually show that in the limit $n\to 0$, the eigenvalues for the longitudinal and anomalous sectors coincide, i.e. 
\be
\Lambda_l=\Lambda_a=-\frac{2(1-\gamma v(\gamma))}{{\cal Q}(\gamma)^2}\leq 0\;,
\ee
which shows that the longitudinal sector is stable as well. As is usually the case, one can thus show that the only instability of the RS ansatz arises from the replicon sector.

\begin{figure}
    \centering
    \includegraphics[width=0.48\linewidth]{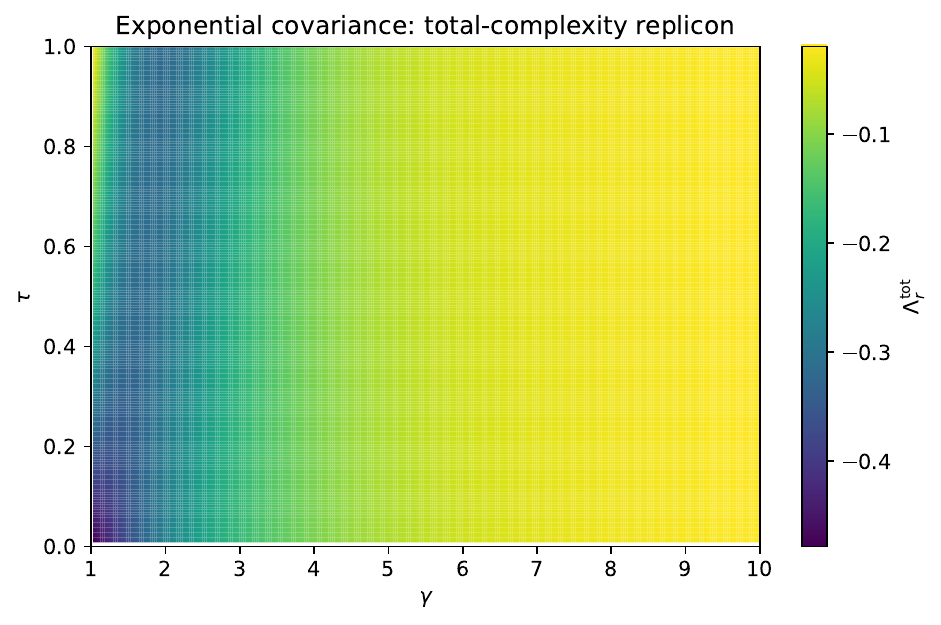}\includegraphics[width=0.48\linewidth]{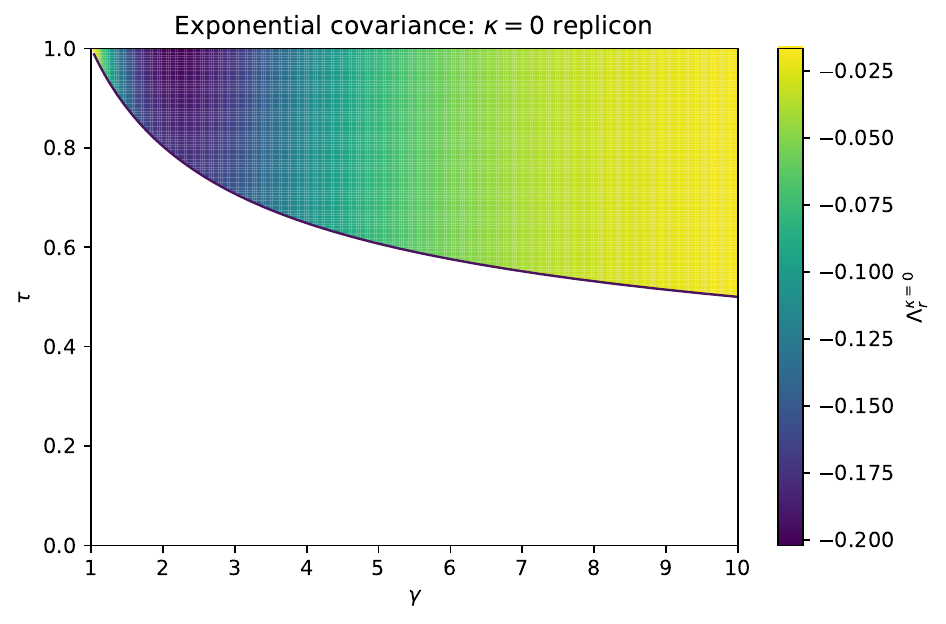}
    \includegraphics[width=0.48\linewidth]{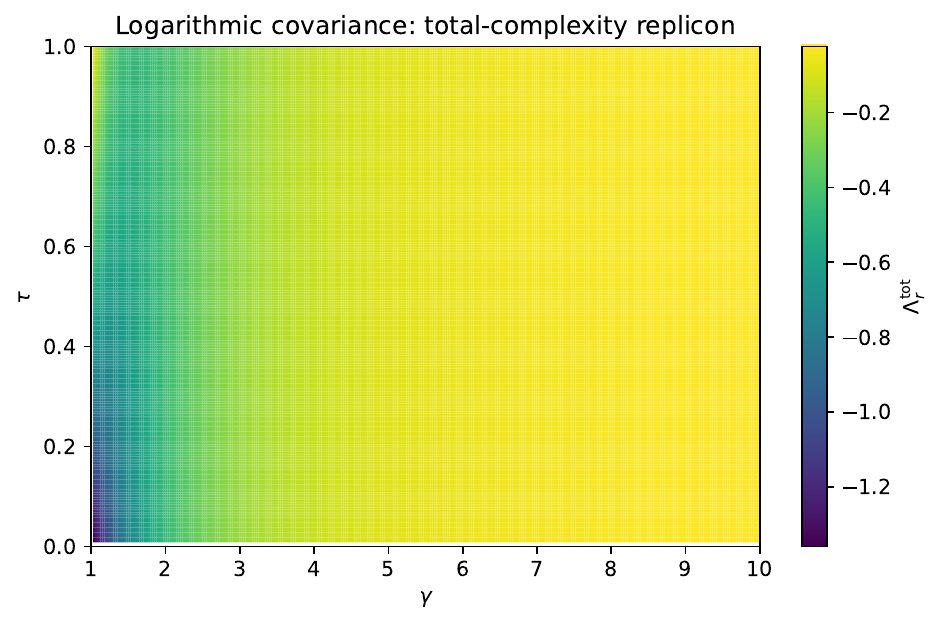}\includegraphics[width=0.48\linewidth]{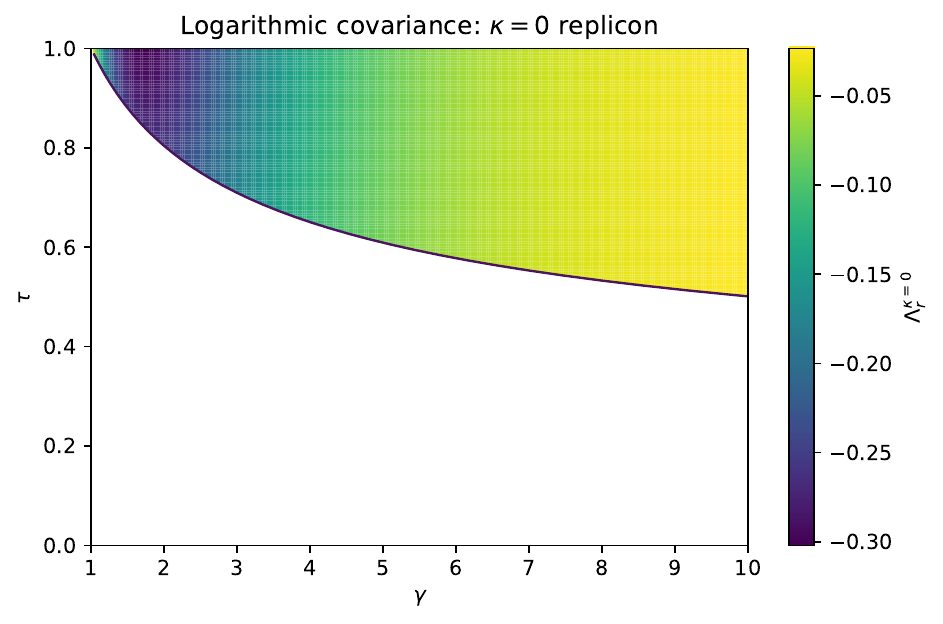}
    \caption{In these figures, we have plotted the replicon eigenvalue $\Lambda_r^{\rm tot}$ for the total complexity and $\Lambda_{r}^{\kappa=0}$ for the complexity at fixed index $\kappa=0$ throughout phase space. The panels on the left correspond to the total complexity while the panels on the right correspond to our prediction for the $\kappa=0$ complexity, which is only valid for $\tau\geq \tau_q$. The top panels correspond to the exponential covariance function $\Gamma(q)=\exp(-q)$ while the bottom panels correspond to $\Gamma(q)=1/\ln(e+q)$. In all the different panels, the replicon eigenvalue is negative throughout the tested phase-space, indicating the stability of the RS ansatz with respect to continuous RSB transitions.}
    \label{fig:replicon}
\end{figure}

To support our analytical results for the quenched complexity, we have performed numerical simulations to test the RS stability and the sign of the replicon in the whole phase-space where the complexity is positive. In Fig. \ref{fig:replicon}, we show the replicon for the total complexity and the complexity of equilibria with fraction $\kappa=0$ for two of the covariances used in the main text, i.e. the exponential covariance $\Gamma(q)=\exp(-q)$ and the covariance $\Gamma(q)=1/\ln(e+q)$. The replicon is negative throughout the full tested phase-space (where the complexity is positive, i.e. for $\tau\geq \tau_q$ in the case $\kappa=0$), showing that the RS ansatz is stable with respect to continuous RSB transition. Note that this however does not rule out a discontinuous RSB transition. This analysis would require considering the 1RSB action and is beyond the scope of this paper.

\section{Factorial moments for $\tau=0$}

Let us finally consider the expression for $\tau=0$, i.e. in the limit where $R(q)\to 0$. In that limit, one obtains that 
\begin{align}
    \Phi_n^{\rm tot}[\mathbb{Q}]=&\frac{1}{2}\min_{L^0,L^1,L^2}\left[{\rm Tr}\left(\sum_{\eta=0}^2 L^{\eta}Q^{\eta}\right)-\int d\nu(\mu)\,\ln \det\left(\sum_{\eta=0}^2\mu^{\eta} \tilde D L^{\eta}\right)\right]\\
    &+\min_{t,C^0,C^1}\left[\frac{1}{2}{\rm Tr}\left(\tilde D P\right)-{\rm Tr}\left(C^1\right)+n\lambda_*(t)t+n\int d^2 z\,\rho_{c,\tau}(z)\,\ln |z-\lambda_*(t)|\right]\;,\\
    P=&\begin{pmatrix}
    C^0&C^1
\end{pmatrix}\mathbb{Q}^{-1}\begin{pmatrix}
    {C^0}^T\\{C^1}^T
\end{pmatrix}\;,\;\;t=-\left.\partial_\lambda \int d^2 z\,\rho_{c,\tau}(z)\,\ln |z-\lambda|\right|_{\lambda=\lambda_*(t)}\;.
\end{align}
The optimisation over $C^0,C^1$ and $t$ yields
\begin{align}
  \tilde D\begin{pmatrix}
   C^0&C^1   
  \end{pmatrix}\mathbb{Q}^{-1}=\begin{pmatrix}
   0&\mathbb{I}   
  \end{pmatrix}\;,\;\;\lambda_*(t)=0\;.  
\end{align}
Multiplying the first matrix equations by  $\mathbb{Q}$, one obtains
\be
C^0=\tilde D^{-1} Q^1\;,\;\;C^1=\tilde D^{-1} Q^2\;.
\ee
Inserting back into the equations, one obtains
\begin{align}
    \Phi_n^{\rm tot}[\mathbb{Q}]=&\frac{1}{2}\min_{L^0,L^1,L^2}{\rm Tr}\left[\sum_{\eta=0}^2 L^{\eta}Q^{\eta}-\int d\nu(\mu)\,\ln\left(\sum_{\eta=0}^2\mu^{\eta} \tilde D L^{\eta}\right)-\tilde D^{-1}Q^2\right]+n\int d^2 z\,\rho_{c,\tau}(z)\,\ln |z|\;.
\end{align}
Optimising with respect to the $L^{\eta}$'s yield
\begin{align}
    Q^{\eta}&=\int d\nu(\mu)\mu^{\eta} \left(\sum_{\omega=0}^2\mu^{\omega} L^\omega\right)^{-1}\;,
\end{align}
such that $\sum_{\eta=0}^2 L^{\eta}Q^{\eta}=\mathbb{I}$ and the final expression before optimising over $\mathbb{Q}$ reads
\be
\Phi_n^{\rm tot}[\mathbb{Q}]=\frac{1}{2}{\rm Tr}\left[\mathbb{I}-\int d\nu(\mu)\,\left(\ln\left(\sum_{\eta=0}^2\mu^{\eta} \tilde D L^{\eta}\right)+\mu^2\left(\sum_{\eta=0}^2\mu^{\eta} \tilde D L^{\eta}\right)^{-1}\right)\right]+n\int d^2 z\,\rho_{c,\tau}(z)\,\ln |z|\;.
\ee
Rewriting $G(\mu)=\left(\sum_{\eta=0}^2\mu^{\eta} \tilde D L^{\eta}\right)^{-1}$, the problem is just a simple optimisation over $G(\mu)$ of the action
\be
\Phi_n^{\rm tot}=\frac{1}{2}\int d\nu(\mu)\,{\rm Tr}\left[\mathbb{I}+\ln G(\mu)-\mu^2 G(\mu)\right]+n\int d^2 z\,\rho_{c,\tau}(z)\,\ln |z|\;.
\ee
It is simply solved by taking $G_*(\mu)=\mathbb{I}/\mu^2$ and yields the simple expression
\be
\lim_{N\to \infty}\frac{1}{N}\ln\moy{\prod_{p=0}^{n-1}\left({\cal N}_{N,\rm tot}-p\right)}=n\Xi_{\rm tot}=n\left[\int d^2 z\,\rho_{c,\tau}(z)\,\ln |z|-\int d\nu(\mu)\ln \mu\right]\;,
\ee
which has been obtained without assumption on the symmetry of the ansatz.

\section{Variance}
We finally use the expression of the factorial moments at arbitrary $n$ to determine the sample-to-sample fluctuations of the number of equilibria. We restrict here to homogeneous relaxation rates $\mu_i=\mu$ and to $k=N\kappa$. 
Specializing ~\eqref{RS_moment} to this case, using the isotropic relations following ~\eqref{q_lambda} for the combinations $\sum_{\eta=0}^2\mu^\eta\lambda_\eta$ and $\sum_{\eta=0}^2\mu^\eta l_\eta$, together with $q_0=q_1/\mu=q_2/\mu^2=q$ and $\Delta_0=\Delta_1/\mu=\Delta_2/\mu^2=\Delta$, one obtains
{\small
\begin{align}
&\lim_{N\to\infty}\frac{1}{N}\ln{\cal M}_{N\kappa}^{n}
=\operatorname*{extr}_{q,\Delta}\Bigg\{
\frac{n}{2}(1+\ln\Delta)
+\frac{1}{2}\ln\left[1+n\left(\frac{q}{\Delta}-1\right)\right]
-\frac{1}{2}\left[
n\ln\left(D(2\Delta)-D(0)\right)
+\ln\left(1-\frac{nD(2\Delta)}{D(2\Delta)-D(0)}\right)
\right]
\nonumber\\
&\qquad
+\frac{n\mu^2}{2}\left[
\frac{q}{D(0)+(n-1)D(2\Delta)}
-\frac{(n-1)\Delta D(2\Delta)}
{[D(2\Delta)-D(0)][D(0)+(n-1)D(2\Delta)]}
+\frac{(n-1)\Delta^2R(2\Delta)}
{[D(2\Delta)-D(0)][D(2\Delta)-D(0)+n\Delta R(2\Delta)]}
\right]
\nonumber\\
&\qquad
+n\left[
\int d^2z\,\rho_{c,\tau}(z)\ln|z-\lambda(\kappa)|
-\frac{
\left(
\lambda(\kappa)+
\frac{(n-1)\mu\Delta R(2\Delta)}
{D(2\Delta)-D(0)+n\Delta R(2\Delta)}
\right)^2
}{
2\left[
R(0)+(n-1)R(2\Delta)
-\frac{n(n-1)\Delta R(2\Delta)^2}
{D(2\Delta)-D(0)+n\Delta R(2\Delta)}
\right]
}
\right]
\Bigg\}
\nonumber\\
&=\operatorname*{extr}_{\Delta}\Bigg\{
\frac{n-1}{2}\left[
1+\ln\frac{\mu^2\Delta}{D(2\Delta)-D(0)}
-\frac{\mu^2\Delta}
{D(2\Delta)-D(0)+n\Delta R(2\Delta)}
\right]
\nonumber\\
&\qquad
+n\left[
\int d^2z\,\rho_{c,\tau}(z)\ln|z-\lambda(\kappa)|
-\ln\mu
-\frac{
\left(
\lambda(\kappa)+
\frac{(n-1)\mu\Delta R(2\Delta)}
{D(2\Delta)-D(0)+n\Delta R(2\Delta)}
\right)^2
}{
2\left[
R(0)+(n-1)R(2\Delta)
-\frac{n(n-1)\Delta R(2\Delta)^2}
{D(2\Delta)-D(0)+n\Delta R(2\Delta)}
\right]
}
\right]
\Bigg\}\;.
\label{moment_homogeneous_n}
\end{align}
}
In the second equality we used the saddle
$q=\frac{n-1}{n}\Delta-\frac{D(0)+(n-1)D(2\Delta)}{n\mu^2}$.

We now expand ~\eqref{moment_homogeneous_n} around $n=0$ as
$\lim_{N\to\infty}N^{-1}\ln{\cal M}_{N\kappa}^{n}
=n\hat \Xi(\kappa)+n^2 {\cal V}(\kappa)/2+O(n^3)$.
Indeed, using $({\cal N})_n=\Gamma({\cal N}+1)/\Gamma({\cal N}-n+1)$, the first two derivatives at $n=0$ give respectively $\moy{\psi({\cal N}+1)}$ and ${\rm Var}[\psi({\cal N}+1)]-\moy{\psi_1({\cal N}+1)}$. Whenever ${\cal N}_{N,N\kappa}$ is exponentially large, $\psi({\cal N}+1)\sim\ln{\cal N}$ while the trigamma contribution is subleading, so that
${\cal V}(\kappa)=\lim_{N\to\infty}N^{-1}{\rm Var}(\ln{\cal N}_{N,N\kappa})$.
It is important at second order to expand the saddle itself,
$\Delta(n)=\Delta+n\,\delta\Delta+O(n^2)$.
At $n=0$ one recovers
$\Delta=[D(2\Delta)-D(0)]/\mu^2$.
The displacement $\delta\Delta$ contributes to the $O(n^2)$ term and cannot in general be neglected. At order $n$, for $\Gamma_V(q)=\tau\Gamma(q)$ and $\Gamma_A(q)=(1-\tau)\Gamma(q)$, one recovers the fixed-index quenched complexity given in ~\eqref{Xi_kappa}.
Keeping the displacement of $\Delta(n)$ at the next order gives
\be
{\cal V}(\kappa)=
\frac{
\left[
\sqrt{\gamma}(1+\tau)z(\kappa)-1-\gamma\tau
\right]^2
}{
\gamma^2\tau^2[1-v(\gamma)]^2
}
\Bigg\{
\gamma\tau v(\gamma)
\left[
1+\gamma\tau v(\gamma)
\right] -\frac{1}{2}
\left[
2\gamma\tau v(\gamma)
-\frac{\gamma v'(\gamma)}{1-v(\gamma)}
\left(
\sqrt{\gamma}(1+\tau)z(\kappa)-1-\gamma\tau
\right)
\right]^2
\Bigg\}\;.
\label{variance_fixed_kappa}
\ee 
Here $v'(\gamma)=dv(\gamma)/d\gamma$
and we recall that $v(\gamma)=\Gamma''({\cal Q}(\gamma))/\Gamma''(0)$, where ${\cal Q}(\gamma)$ is determined implicitly by $\gamma={\cal Q}(\gamma)\Gamma''(0)/[\Gamma'({\cal Q}(\gamma))-\Gamma'(0)]$. Equivalently, at the $n=0$ saddle, $\Delta={\cal Q}(\gamma)/2$ and $R(2\Delta)=R(0)v(\gamma)=\mu^2\gamma\tau v(\gamma)$. 

For the total number of equilibria, the extremisation over the common RS index gives
$z(\kappa_*)=[1+\tau\gamma v(\gamma)]/[\sqrt{\gamma}(1+\tau v(\gamma))]$.
The $O(n)$ displacement of $\kappa_*$ does not contribute at order $n^2$ since $\Xi'(\kappa_*)=0$, and hence ${\cal V}_{\rm tot}={\cal V}(\kappa_*)$. One thus finds
\be
{\cal V}_{\rm tot}
\equiv
\lim_{N\to\infty}
\frac{{\rm Var}(\ln{\cal N}_{\rm tot})}{N}
=
\frac{
(\gamma-1)^2
}{
\gamma^2[1+\tau v(\gamma)]^2
}
\Bigg\{
\gamma\tau v(\gamma)
\left[
1+\gamma\tau v(\gamma)
\right]
-\frac{\gamma^2\tau^2}{2}
\left[
2v(\gamma)
+\frac{
(\gamma-1)v'(\gamma)
}{
1+\tau v(\gamma)
}
\right]^2
\Bigg\}\;.
\label{variance_total}
\ee 

\begin{figure}
    \centering
    \includegraphics[width=0.48\linewidth]{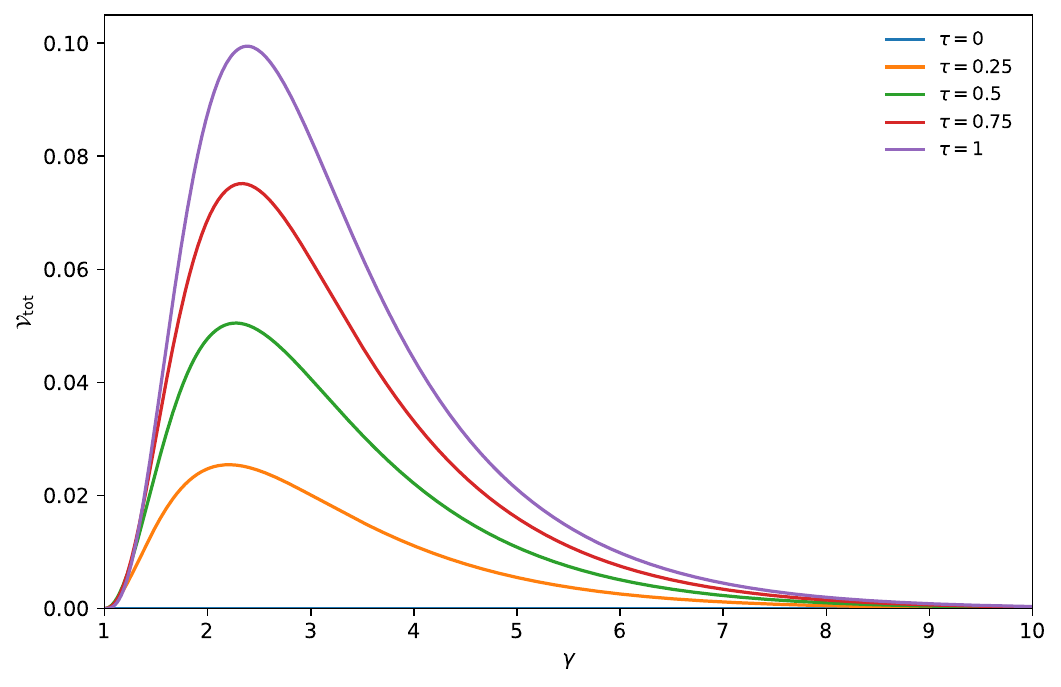}
    \includegraphics[width=0.48\linewidth]{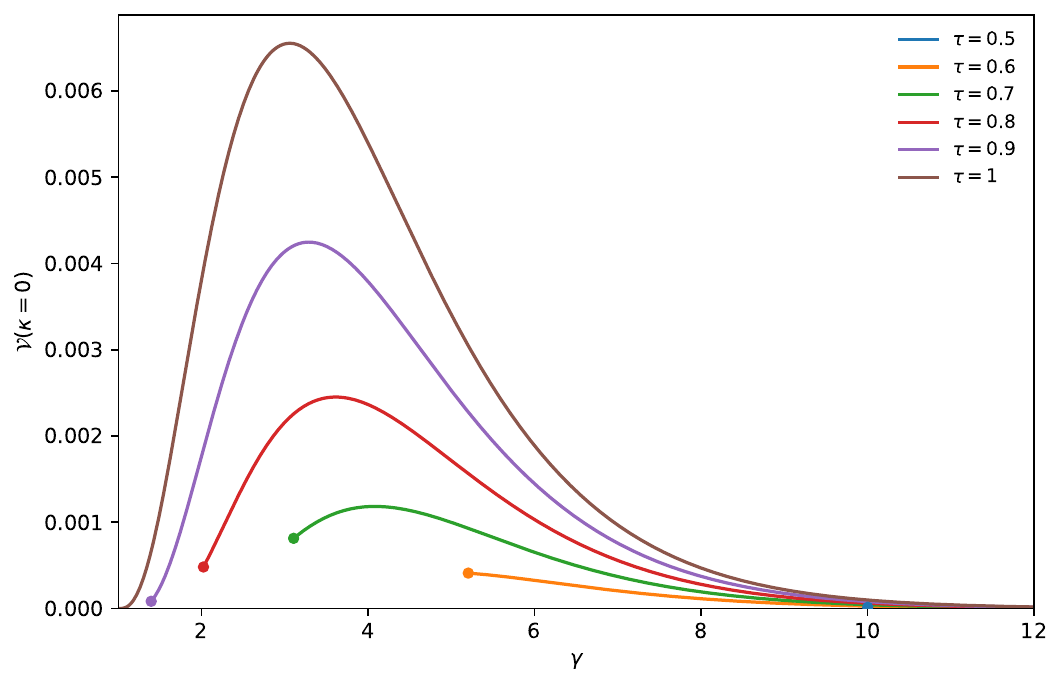}
    \caption{Plot of the total rescaled variance ${\cal V}_{\rm tot}$ (left panel) and ${\cal V}(\kappa=0)$ (right panel) as a function of $\gamma$ for several values of $\tau$. The plot is done for the exponential covariance $\Gamma(q)=\exp(-q)$. Interestingly, at the transition $\tau\to\tau_q(\gamma)$, which corresponds to the dots on the right panel, the quenched complexity (i.e. the average logarithm of the number of equilibria) $\Xi(0)=\lim_{N\to \infty}\frac{1}{N}\moy{\ln({\cal N}_{N,0}+1)}$ vanishes but its variance ${\cal V}(\kappa=0)>0$ remains finite.}
    \label{fig:variance}
\end{figure}

Close to the trivialisation transition one has
$v(\gamma)=1-2(\gamma-1)+O((\gamma-1)^2)$, therefore
\begin{align}
{\cal V}_{\rm tot}
=&
\frac{\tau(1-\tau)}{(1+\tau)^2}
(\gamma-1)^2
+\frac{3\tau(3\tau-1)}{(1+\tau)^3}
(\gamma-1)^3
+O((\gamma-1)^4)\;.
\label{variance_transition}
\end{align}
For $0<\tau<1$ the extensive variance thus turns on quadratically at $\gamma=1$, whereas for the gradient case $\tau=1$ the quadratic contribution vanishes and
${\cal V}_{\rm tot}=\frac34(\gamma-1)^3+O((\gamma-1)^4)$.
In the fully non-reciprocal limit $\tau=0$, ~\eqref{variance_total} gives ${\cal V}_{\rm tot}=0$ for all $\gamma>1$: fluctuations of $\ln{\cal N}_{N,\rm tot}$ are then sub-extensive. At fixed $\kappa$, the interpretation of~\eqref{variance_fixed_kappa} as a variance of $\ln{\cal N}_{N,N\kappa}$ requires the corresponding index sector to contain exponentially many equilibria. A plot of the rescaled variance is shown in Fig. \ref{fig:variance}, showing positivity for any $\gamma> 1$ and $\tau>0$. In the trivial phase $\gamma<1$ and for any $0\leq \tau\leq 1$, it was shown in \cite{ben2021counting} that ${\cal N}_{N,\rm tot}={\cal N}_{N,0}=1$ with probability tending to one as $N\to \infty$. At fixed $\kappa$, the probability of having exactly zero equilibria can be bounded in terms of the average and variance of $\psi({\cal N}_{N,N\kappa}+1)$, i.e.
\be
{\rm Prob}.\left[{\cal N}_{N,N\kappa}=0\right]\leq \frac{{\rm Var}\left[\psi({\cal N}_{N,N\kappa}+1)\right]}{\left(\psi(1)-\moy{\psi({\cal N}_{N,N\kappa}+1)}\right)^2}\;.
\ee
Thus, when both ${\cal V}(\kappa)=\lim_{N\to \infty}{\rm Var}\left[\psi({\cal N}_{N,N\kappa}+1)\right]/N=O(1)$ and $\hat \Xi(\kappa)=\lim_{N\to \infty}\moy{\psi({\cal N}_{N,N\kappa}+1)}/N=O(1)$ and positive, one obtains that the probability to have zero equilibria goes to zero identically $\lim_{N\to \infty} {\rm Prob}.\left[{\cal N}_{N,N\kappa}=0\right]=0$. This identity holds throughout the phase where $\hat \Xi(\kappa)=\Xi(\kappa)>0$. Conversely, in the phase where the annealed complexity is negative $0>\Sigma(\kappa)\geq\lim_{N\to \infty}\frac{1}{N}\ln {\rm Prob}.\left[{\cal N}_{N,N\kappa}>0\right]$, the number ${\cal N}_{N,N\kappa}$ is zero with probability tending to one as $N\to \infty$.

\section{Mechanism favoring unstable fixed points in the annealed counting}
Comparing the results from the annealed and quenched complexity, the equilibria dominating the average count are less stable than typical ones. In fact, a simple mechanism ensures that the annealed calculation overestimates not only the total number of equilibria but also their instability, and that this effect becomes stronger as $\tau$ increases. This mechanism is rather generic: it relies on the presence of a confining potential, and on the fact that the random forces have translationally invariant statistics, and admit the decomposition into a potential and a divergence-free term. For ease of notation, we illustrate the argument in the case $\Gamma_V=\tau\Gamma$, $\Gamma_A=(1-\tau)\Gamma$ and for homogeneous rates $\mu_i=\mu$, and we stick to the rescaled variables of Sec. \ref{sec:RKR}. 

Consider the shifted Jacobian \eqref{prob_jac} at an equilibrium $\mathbf{x}_a$, conditioned on the $\mathbf{x}_b$ with $b=1, \cdots, n$ being fixed points. Taking the trace of \eqref{prob_jac}  and dividing by $N$, one has $N^{-1}\mathrm{Tr}\,X^a=O(1/N)$, since the elliptic bulk is centred, and $N^{-1}\mathrm{Tr}\,W^a=O(1/N)$, since $W^a$ has finite rank. Hence, to leading order in $N$, $\xi_a= N^{-1}\mathrm{Tr}\,[J(\mathbf{x}_a)]=N^{-1}\nabla\cdot\mathbf{F}(\mathbf{x}_a)$. Since the non-reciprocal part of the force is divergence-free, to leading order in $N$, then $\xi_a=-N^{-1}\Delta V(\mathbf{x}_a)$: the shift $\xi_a$ is therefore the divergence of the gradient part of the force. As discussed in Materials and Methods, the fluctuations of $\xi_a$ control the extensive index of the equilibria: at exponential order, the larger $\xi_a$, the more unstable the equilibrium. We now argue that the number of equilibria and their instability are positively correlated. Indeed, consider a set of $n$ fixed points $\mathbf{x}_b$, satisfying $F_i(\mathbf{x}_b)=\mu x_{bi}$. The statistics of the $\xi_a$, conditioned on all the $\mathbf{x}_b$ being fixed points, 
is given within the RS ansatz in \eqref{av1} and \eqref{av2}. The mean is the same for all $a$, and the covariance has the form $N\,\mathrm{Cov}(\xi_a,\xi_b)=\alpha_n\,\delta_{ab}+\beta_n$, where $\alpha_n$ and $\beta_n$ can be reconstructed from \eqref{av2}. All these terms depend on the functions $D(q)=2\Gamma'(q)$ and $R(q)=4\tau\,\Gamma''(q)$ that are, to leading order in $N$, the covariances of the forces and of the unconditioned shifts $\xi_a$. In particular, $R(0)=c\tau$. Within the RS solution, the typical distance  $d^2_{ab}=N^{-1}|\mathbf{x}_a-\mathbf{x}_b|^2$ between equilibria is fixed by the saddle point, $d^2_{ab}=2\Delta_0=\mathcal{Q}(\gamma)$ for $a\neq b$. Moreover, $N\Delta_1=N \mu\, \Delta_0=\mu\,\mathbf{x}_b\cdot(\mathbf{x}_b-\mathbf{x}_a)$ is the projection of the force at the fixed point $\mathbf{x}_b$ onto the segment joining it to $\mathbf{x}_a$. The crucial observation is that this quantity, that appears in the average shift \eqref{av1}, is positive. Its positivity has a simple origin: it only relies on the confining term being the gradient of a convex potential, and it holds within the RS formalism.

Now, the annealed calculation of the complexity is an average over a size-biased ensemble: it weights each realization of the disorder by its number of fixed points, and therefore favors realizations with many of them. Because of the positive correlations, realizations with more fixed points also have a larger average divergence $\bar\xi$: their fixed points are on average less stable. This can be seen even more explicitly. Since all pairs of conditioned shifts $\xi_a$ have the same covariance, they can be written as
\begin{equation}
\xi_a=\eta+\zeta_a,\qquad \langle\eta\rangle=\bar\xi,\quad \mathrm{Var}(\eta)=\frac{\beta_n}{N},\qquad \langle\zeta_a\rangle=0,\quad \mathrm{Var}(\zeta_a)=\frac{\alpha_n}{N},
\end{equation}
with $\eta$ common to all fixed points and the $\zeta_a$ independent of each other and of $\eta$. We remind that we are using the notation $N\,\mathrm{Cov}(\xi_a,\xi_b)=\alpha_n\,\delta_{ab}+\beta_n$. The shared part $\eta$ is a property of the realization of the disorder, not of any single fixed point: when $\xi_a$ is averaged over the exponentially many fixed points of a given realization, the individual parts $\zeta_a$ average out and only $\eta$ remains. In other words, $\eta$ can be interpreted as the divergence of the force field felt by the whole population of fixed points of a realization, while $\zeta_a$ is the local deviation at each of them. Given that $\eta$ shifts all the Jacobian spectra of a realization at once, a realization with a larger $\eta$ has less stable fixed points. These are precisely the realizations selected by the annealed average. More generally, in the replica calculation of the $n$-th moment, each extra factor of $\mathcal{N}_{N,k}$ in the weight appears as one more replica conditioned to be a fixed point, and pushes $\eta$ further towards positive values, enhancing this effect. In the quenched calculation ($n\to0$) this weighting is removed, and the mean of $\eta$ is shifted with respect to the annealed one by $\bar\xi|_{n\to0}=-R(2\Delta_0)\Delta_1/[D(2\Delta_0)-D(0)]<0$, see \eqref{av1}. The realizations dominating the average therefore have both more fixed points and a larger shared divergence, i.e.\ less stable fixed points, than typical ones. This mechanism is ineffective for $\tau=0$, where the bias $\bar\xi$ vanishes; consistently, in this limit we find that quenched and annealed counting coincide.

This mechanism allows us  to interpret our results quantitatively, and not only in terms of trends. Imposing an index $k=N\kappa$ on all replicas requires $\xi_a=-\lambda(\kappa)$ for every $a$, i.e. it fixes all shifts to the same value. The probability of this event depends only on the mean $\bar\xi$ of the shifts and on the fluctuations of their sum $\sum_a\xi_a$, whose variance (per replica) is $\alpha_n+n\beta_n$; see \eqref{eq:IndexFix}. In the annealed calculation ($n=1$), the shifts have zero mean and the variance equals $R(0)=c\tau$: both the shared and the individual parts can be tuned to reach a given index. In the quenched calculation ($n\to0$), the mean is shifted by a fixed, negative amount, and the variance reduces to $\alpha_0=R(0)-R(2\Delta_0)=c\tau(1-v(\gamma))$, with $v(\gamma)=R(\mathcal{Q}(\gamma))/R(0)$: only the individual parts can be tuned, since the shared part is fixed by the realization of the disorder, which has to be typical. These two modifications have distinct consequences. First, since a lower mean of the shift makes it cheaper to reach larger values of $\lambda(\kappa)$, i.e.\ smaller values of $\kappa$, the index $\kappa_q$ dominating the quenched complexity is smaller than its annealed counterpart $\kappa_a$: typical fixed points are more stable. Second, the reduced variance makes deviations from the dominant index more costly, so that the quenched complexity $\Xi(\kappa)$ is both smaller than and more sharply peaked around its maximum than the annealed one; this is the origin of the denominator $2\gamma\tau(1-v(\gamma))$ appearing in the expression of $\Xi(\kappa)$. Both effects are controlled by the same quantity, $R(2\Delta_0)=c\tau v(\gamma)$, which sets both the shift of the mean and the variance lost in going from the annealed to the quenched calculation: it grows with $\tau$, and vanishes for $\tau=0$ or $v(\gamma)\to0$. This explains why the discrepancy between quenched and annealed counting disappears at $\tau=0$, as well as for $\gamma\to\infty$, where $v(\gamma)\to0$, as shown in the figures of the main text.